\documentclass[aps,prl,floatfix,showpacs,twocolumn,amsmath,amssymb]{revtex4-2}

\usepackage[caption=false]{subfig}
\usepackage{graphicx}
\usepackage{bm}
\usepackage{booktabs}

\usepackage{hyperref}
\usepackage{color}
\usepackage{amssymb}
\usepackage{epstopdf}
\usepackage{upgreek}
\usepackage{comment}
\usepackage{todonotes}

\begin{document}

\title{Gouy phase-matched angular and radial mode conversion in four-wave mixing}

\author{Rachel F.\ Offer}
\email{rachel.offer@adelaide.edu.au}
\affiliation{Department of Physics, SUPA, University of Strathclyde, Glasgow G4 0NG, UK}
\author{Andrew Daffurn}
\affiliation{Department of Physics, SUPA, University of Strathclyde, Glasgow G4 0NG, UK}
\author{Erling Riis}
\affiliation{Department of Physics, SUPA, University of Strathclyde, Glasgow G4 0NG, UK}
\author{Paul F. Griffin}
\affiliation{Department of Physics, SUPA, University of Strathclyde, Glasgow G4 0NG, UK}
\author{Sonja Franke-Arnold}
\affiliation{School of Physics and Astronomy, SUPA, University of Glasgow, Glasgow G12 8QQ, UK}
\author{Aidan S.\ Arnold}
\email{aidan.arnold@strath.ac.uk}
\affiliation{Department of Physics, SUPA, University of Strathclyde, Glasgow G4 0NG, UK}

\date{\today}

\begin{abstract}


Studying the conversion between transverse light modes via four-wave mixing in a heated rubidium vapour, we demonstrate and explain a transfer between azimuthal and radial mode numbers. These  relate to orthogonal modal dimensions, which one would not normally expect to interact.  While angular momentum conservation in this nonlinear process dictates the selection rules for the angular mode number, the role of the radial mode number is more esoteric. 
We demonstrate systematically that the Gouy phase is the key to understanding this conversion, leading to strikingly different conversion behaviour in the thick and thin medium regime. 
Our experimental investigation of the transition between these regimes bridges the gap between previous experiments in atomic thick media and work in nonlinear crystals. Our work sets a clear starting point to explore new territory in the thick medium regime, allowing efficient radial-to-azimuthal and radial-to-radial mode conversion.

\end{abstract}
\maketitle

Laguerre-Gauss (LG) beams and their corresponding orbital angular momentum (OAM) have been a burgeoning research area since their association in 1992  \cite{Allen1992}. They have applications in microfabrication \cite{Ni2017}, entanglement protocols \cite{Pan2019,Ding2016}, 
multiplexing in classical as well as quantum communication systems \cite{Gibson2004,Yue2018,Bozinovic2013}, and are summarised in a plethora of reviews and roadmaps  \cite{Yao2011rev,Bliokh2015,Rubinsztein2017,FrankeArnold2017}.
These studies have been devoted almost exclusively to the study of the azimuthal mode index $\ell$ of the LG modes, while the radial index $p$ remains largely unexplored. 

The azimuthal mode number has a simple geometrical interpretation:  beams with an azimuthal phase winding $\exp(i \ell \theta)$ carry an OAM of $\ell \hbar$ per photon, a quantum interpretation of the Fourier conjugate relation between angle and angular momentum. 
Understanding the radial mode number however is less intuitive, as the observed $p$-mode decomposition of a beam depends on the chosen fundamental waist, and a generating function is not readily identified  \cite{Karimi2012}. Nevertheless various strategies have been devised to assess radial modes \cite{Dholakia2012,Dudley2013,Zhou2016APL,Fontaine2019,Boyd2017,Zeilinger2018}, and several recent publications highlight their role in detailed quantum investigations  \cite{Loeffler2012,Boyd2014HOM,Boyd2014,Krenn2015,Zhang2018,Boyd2019}.

In nonlinear processes, including spontaneous parametric down-conversion (SPDC), second harmonic generation (SHG) and four-wave mixing (FWM), the full 3D overlap of all participating modes determines the efficiency of mode conversion \cite{Yao2011,Shao2013}. The transverse overlap, specifically the integration over the azimuthal angle, is sufficient to define phase-matching, and hence angular momentum conservation in nonlinear crystals \cite{Courtial1997,Mair2001, Franke-arnold2002b,Shao2013} and in atomic vapours \cite{Walker2012,Akulshin2016,Chopinaud2018,Offer2018}.  
In extended media however, 
the relative Gouy phase of the involved modes becomes important \cite{Khoury2017,Khoury2018,Khoury2019,Zhu2020,Khoury2020,Zhao2020}.

Here, we perform and analyse FWM of arbitrary LG modes in an atomic vapour,
reaching, with the same experimental setup, the equivalent of both the thin and thick crystal limit which are characterized by strikingly different behaviour. This allows us to investigate nonlinear interaction regions beyond the Rayleigh range, a regime not usually accessed in crystal-based SPDC and SHG.
We explain the difference in behaviour by a phase-matching condition -- Gouy phase matching -- which gains importance as the interaction length reaches and exceeds the Rayleigh range.  
In the thick medium regime this restricts the number of modes generated, allowing us to demonstrate clean novel azimuthal-to-radial and radial-to-radial mode conversion. 

Conversion between modes of different azimuthal mode number has been associated with angular momentum conservation \cite{Mair2001, Franke-arnold2002b}, however there exists no obvious conservation law that would lead to a conversion between radial and azimuthal mode number. Coupling between orthogonal mode indices can result from symmetry breaking via dissipation or boundary conditions and has been observed on platforms as diverse as optical fibers and waveguides, acoustics, stellar interiors and supercooled liquids. These effects are usually seen as detrimental and lead to a loss of purity. This contrasts our observations for mode conversion via four-wave-mixing in the thick medium regime experiments, where Gouy phase matching facilitates efficient and controlled mode manipulation.

\begin{figure}[!t]
\centering
\includegraphics[width=\columnwidth]{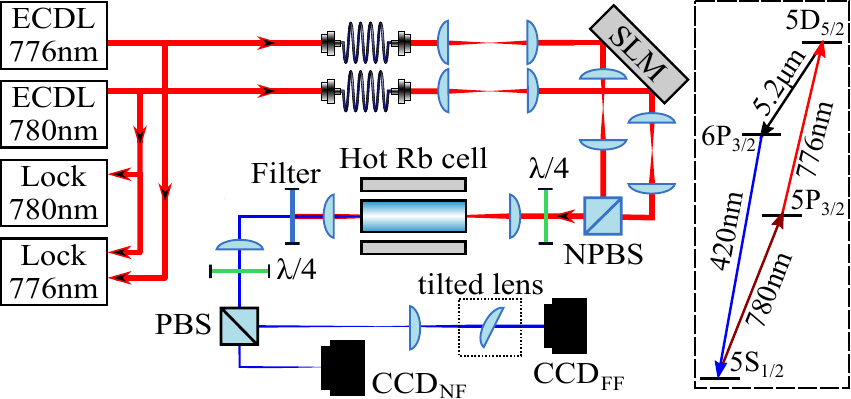}
\caption{Experimental setup and level scheme. Fibre-coupled $780\,$nm and $776\,$nm pump beams from extended-cavity diode lasers (ECDLs), are independently shaped on an SLM  (incidence angle exaggerated) into pure LG$_p^{\ell}$ modes. They are relayed through beam telescopes, combined on a non-polarising beamsplitter (NPBS), co-circularly polarised with a quarter-wave plate ($\lambda/4$) and focused into a hot rubidium vapor cell. A filter isolates the blue light generated by FWM. The near- and far-field from the Rb cell is imaged on cameras CCD$_\textrm{NF}$  and CCD$_\textrm{FF}$, and analysed using a removable tilted lens.}
\label{fig:setup}
\end{figure}

Before discussing the experiments we briefly outline our theoretical model.
We predict FWM efficiency by evaluating the mode overlap integral of all involved complex electric fields, following Refs~\cite{Walker2012,Lanning2017,Offer2018}, 
\begin{equation}
\int_0^{2\pi}d\theta \int_0^{\infty}r\, dr \int_{-L/2}^{L/2}dz \,u_{780}\, u_{776}\, u^\ast_{\mathrm{IR}}\, u^\ast_{\mathrm{B}},  
\label{eq:overlap}
\end{equation}
where the subscripts denote the input electric fields at wavelengths 780$\,$nm and 776$\,$nm, and the generated fields at 5233$\,$nm (IR) and 420$\,$nm (B) (Fig.~\ref{fig:setup} inset). We use cylindrical polar coordinates, and assume that the pump fields are focused at the centre of a vapour cell, length $L$. 

Any paraxial beam $u(r,z,\theta)$ can be decomposed in the LG$_p^{\ell}$ basis (for more detail see Supplementary Section S1), 
\begin{eqnarray}
    \textrm{LG}^{\ell}_p(r,z,\theta)& =& A^{|\ell|}_p(r,z)\;\,e^{i \ell \theta}\;\,e^{i \Phi_\textrm{G}(z)} , 
    \label{eq:lgmode}
\end{eqnarray}
where $A^{|\ell|}_p$ is a complex amplitude and the azimuthal phase is $\ell \theta$.  The Gouy phase 
\begin{equation}
  \Phi_\textrm{G}(z)= -(1+2p+|l|)\arctan(z/z_\textrm{R}),  
\end{equation}
describes the phase evolution of a mode propagating through a focus, with $z_\textrm{R}=\pi w_0^2/\lambda$ denoting the Rayleigh range for a beam waist $w_0$. The relative Gouy phase between modes with different mode number $N_p^\ell=1+2p+|l|$ crucially affects mode conversion. 
 
The azimuthal integral of Eq.~\ref{eq:overlap} leads to the condition that OAM must be conserved
\begin{equation}
\ell_{780}+\ell_{776}=\ell_{\mathrm{IR}}+\ell_{\mathrm{B}}.  \label{eq:bsoam}
\end{equation} 

We assume the Boyd criterion \cite{Boyd1968} holds, i.e.\ the waists of the generated fields yield the same Rayleigh range as the pump fields. Evaluating the axial integral introduces a dependence on the relative Gouy phase of the modes. Fields with differing mode number $N_p^\ell$ slip out of phase as they propagate through the cell, leading to a reduced FWM efficiency.

In the extreme thick crystal limit, where $z_{\mathrm{R}}\ll L$, the Gouy phase of each field changes by $N^{\ell}_p\pi$ as it propagates through the cell.  
For the beams to remain (Gouy) phase-matched, the total mode order must therefore also be conserved, leading to the condition 
\begin{equation}
(N_p^\ell)_{780}+(N_p^\ell)_{776} = (N_p^\ell)_\mathrm{IR}+(N_p^\ell)_{B}.
\label{eq:bsgouy}
\end{equation}
Only mode combinations that obey Eqns.~\ref{eq:bsoam} and \ref{eq:bsgouy} contribute to FWM.  This is strictly true only in the thick crystal limit.  For any extended medium, however, the evaluation of the axial mode overlap integral results in a decreased efficiency of mode combinations with unmatched Gouy phases.


In our experiment the $420\,$nm light is generated as a superposition of one or more modes, depending on the ratio of cell length to Rayleigh range.  The presence of multiple coherent modes becomes apparent from the spatial variation of the light's intensity on propagation --  individual LG modes, and their incoherent mixtures, have a self-similar intensity profile upon propagation, with radial size scaling as the beam radius. Interference between coherent superpositions with differing mode number and hence Gouy phase, however, results in a modulation of the intensity profile \cite{Paterson2001,Amico2005,Franke-Arnold2007a,Bhattacharya2007,Arnold2012}, with notable differences between near and far field profiles. 
This behaviour serves as a simple way of determining mode population and coherence experimentally.  We additionally analyse the modal decomposition in a single plane using a tilted lens technique \cite{Siemens2016}. 


Our FWM scheme (Fig.~\ref{fig:setup}) utilises the extreme single-pass efficiencies found even at milliWatt powers in the highly non-degenerate diamond atomic level system of alkali metal vapours \cite{Zibrov2002,Meijer2006,Schultz2009,Akulshin2009,Vernier2010,Offer2016,Brekke2016}. 
Near-infrared external cavity diode lasers (ECDLs \cite{Arnold1998}) at $780\,$nm and $776\,$nm pump the input transitions of an $^{85}$Rb diamond level system (inset Fig.~\ref{fig:setup}), with laser and lock details given in the Supplementary Section S2.
FWM, seeded by initial spontaneous emission, generates coherent light beams at $5.2\,\upmu$m and $420\,$nm, in the cascade back to the ground state. 
We use two separate regions of the same SLM to independently shape the complex amplitudes of the $780\,$nm and $776\,$nm pump laser beams, generating high purity LG modes  \cite{Clark2016,Offer2018}. 

Our experimental setup permits us to vary smoothly between a `thick' ($z_\textrm{R} \ll L$)  and `thin' medium ($z_\textrm{R} \geq L$), allowing us to reconcile the respective selection rules for mode conversion in FWM in the different regimes. 
We adjust the beam waist $w_0$ of the pump lasers by using different telescopic optics, thereby changing the Rayleigh length $z_\textrm{R}$ relative to the fixed cell length of $L=25\,$mm.

\begin{figure*}[!t]
\centering
\begin{minipage}{.69\textwidth}
\includegraphics[width=\columnwidth]{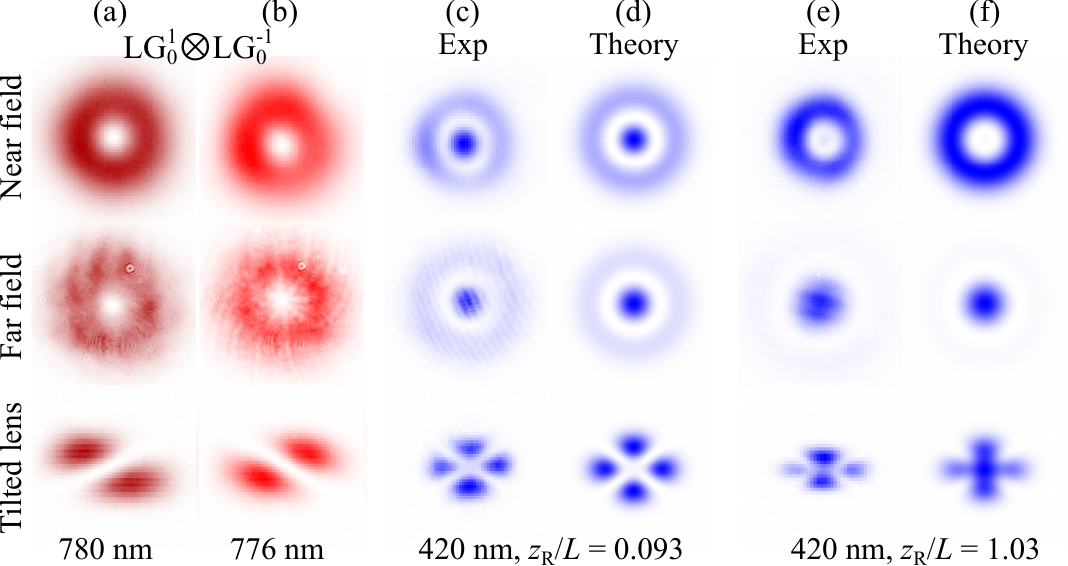}
\end{minipage}
\begin{minipage}{.3\textwidth}
\includegraphics[width=\columnwidth]{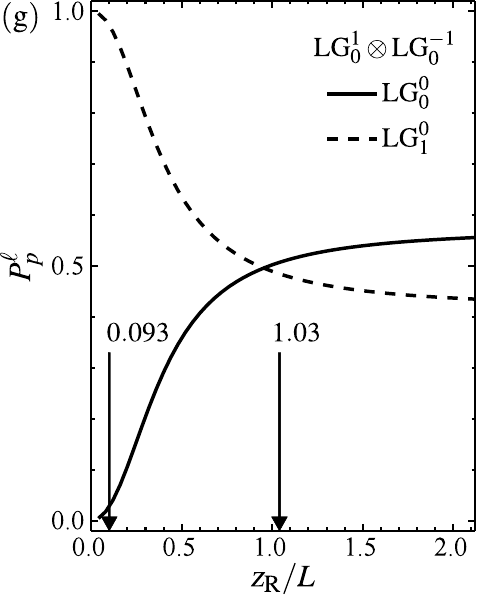}
\end{minipage}
\caption{FWM for pump modes with opposite $\ell$: LG$_{0}^{1}$ at 780$\,$nm (a) and LG$_{0}^{-1}$ at 776$\,$nm (b). 
For a thick medium, we observe $420\,$nm light in an almost pure LG$_1^0$ mode (c), in agreement with our model (d). For a thin medium the experimental (e) and predicted (f) light is a coherent superposition of LG$_0^0$ and LG$_1^0$. 
Each image triplet in (a)-(f) corresponds to the near-field (top) far-field (middle) and tilted lens (bottom) beam intensity. Individual images are peak normalised, and colour saturation corresponds linearly to light intensity.
In (g) we simulate the impact of the medium thickness on the mode purity of the generated light, with  LG$_0^1\otimes \textrm{LG}_1^{-1}$ input. 
}
\label{fig2:adder}
\end{figure*}

We first investigate the conversion of pump modes with opposite angular mode indices to FWM light with a non-zero radial mode index, and examine the behaviour in the thick and thin medium regimes. 
Specifically, we demonstrate FWM for the case of 
$u_{780}=\textrm{LG}^1_0$ and $u_{776}=\textrm{LG}^{-1}_0$ pump light. Fig.~\ref{fig2:adder}(a) and (b) depict the corresponding experimental beam profiles observed in the near- and far-field as well as through a tilted lens \cite{Siemens2016} to visualise their angular mode number $\ell$, where the number of lobes is given by $|\ell|+1$ and the tilt corresponds to the sign of the OAM. 

The $5.2\,\upmu$m infrared light is currently absorbed by our glass cell, but could be readily detected with a sapphire cell. 
If the unobserved IR beam is dominantly generated in the mode $u_\textrm{IR}=\textrm{LG}^0_0$, OAM conservation (Eq.~\ref{eq:bsoam}) dictates that $\ell_{\mathrm{B}}=0$. The only solution also satisfying Gouy phase matching (Eq.~\ref{eq:bsgouy}) is $p_{\mathrm{B}}=1$, thus in the thick crystal limit we expect $u_\textrm{B}=\textrm{LG}^0_1$. This is a counter-intuitive result, as it allows the generation of light with on-axis intensity from modes with a central vortex with associated null intensity. 


The resulting FWM light at $420\,$nm is shown for the thick and thin medium limit in Fig.~\ref{fig2:adder} (c) and (e), respectively. The thick (thin) regime is realised by setting the pump waist to $w_0=24\,\upmu$m ($80\,\upmu$m), corresponding to an `inverse thickness' parameter $z_\textrm{R}/L=0.093$ ($1.03$). 
In the thick medium regime, Fig.~\ref{fig2:adder} (c), the LG$_{0}^{1}$ and LG$_{0}^{-1}$ pump light generates an almost pure blue LG$_{1}^{0}$ mode, easily recognised by its bright central spot and surrounding ring, separated by a ring of zero intensity. Moreover, the beam is self-similar in both the near (top image) and far field (centre), illustrating its single mode nature. The tilted lens image (bottom) corroborates this interpretation, with $\ell_\textrm{B}=0$ as determined by the difference between the number of lobes in the diagonal and anti-diagonal directions, and $p_\textrm{B}=1$ by the minimum of these numbers minus 1. These results agree with a simulation of the beam profiles based on evaluating the mode overlap integral Eq.~\ref{eq:overlap} shown in Fig.~\ref{fig2:adder} (d).

The situation is markedly different in the thin medium regime, where Gouy phase-matching is relaxed, while OAM conservation holds strictly. Evaluating the mode overlap integral shows that the beam is generated in an almost equal superposition of LG$_0^0$ and LG$_1^0$. This is confirmed by our observations and simulations in Fig.~\ref{fig2:adder} (e) and (f).  The near field is a lone intensity ring reminiscent of the pump field overlap, and the far field is dominated by a central spot. This arises because the differing mode numbers, $N_0^0=1$ and $N_1^0=3$ modify the modal interference from the near to the far field, yielding a $\pi$ Gouy phase shift. Moreover, the tilted lens image confirms that $\ell=0$ and we no longer have a single mode, as the spots have no clear nodal delineation.  Further experimental results for $0.49\leq z_\mathrm{R}/L \leq 1.03$ are included in Supplementary Fig.~S1. 

We support our observations by evaluating the 3D overlap integral of Eq.~\ref{eq:overlap} as a function of medium thickness, identifying the relative efficiency of all possible output mode combinations $u_\textrm{B} u_\textrm{IR}$. We note that this simple model does not include the effect of the atoms on the optical propagation, e.g.~via absorption, Kerr lensing or stimulated processes.  In the thick medium regime this results in the generation of all mode pairs obeying OAM conservation and Gouy phase matching, i.e.~LG$_1^0$ LG$_0^0$, LG$_0^0$ LG$_1^0$, LG$_0^1$ LG$_0^{-1}$, LG$_0^{-1}$ LG$_0^{1}$ in the ratio 0.36:0.36:0.13:0.13 and negligible power in a variety of additional modes, with more modes becoming viable for thinner media.

Interestingly our simulations agree best with experimental observations if we constrain the infrared to LG$_0^0$, regardless of medium thickness.  In our previous work with LG$_{p=0}^{\ell>0}$ input modes \cite{Walker2012, Offer2018} this arose naturally for low $\ell$ as the mid-infrared light has a larger waist from the Boyd criterion.  We note that the situation for modes with $p>0$ is more complex as their size no longer scales as a direct function of the mode index.  Instead, we expect that the selection of the infrared LG$_0^0$ mode is a consequence of the full propagation dynamics. Fig.~\ref{fig2:adder} (g) shows the predicted mode decomposition assuming $u_\textrm{IR}=\textrm{LG}^0_0$.  This assumption is applied to all theoretical results in this work.  

We compare our work with earlier investigations on mode conversion. While we show results in the thick and thin medium regimes with $z_\textrm{R}/L \simeq 0.1$ or $1$, respectively, experiments performed using SHG \cite{Khoury2020, Khoury2017,Zhu2020} appear to all have $z_\textrm{R}/L \geq 10$, at least an order of magnitude thinner than demonstrated here. The observed second harmonic beam profile in these publications agrees with our assumption that in this thin regime total mode order conservation is relaxed, so that a spectrum of radial modes can be generated. According to our model, the most drastic changes to mode conversion happen for $z_\textrm{R}/L \leq 2$, before reaching an asymptotic limit.
Additionally, in a FWM system similar to our own, but with fixed $z_\textrm{R}\sim L$ \cite{Akulshin2016}, the $420\,$nm light was observed as a ring-shaped intensity profile with slight on-axis intensity, indicative of partial conversion to radial modes.

\begin{figure}[!t]
\centering
\includegraphics[width=\columnwidth]{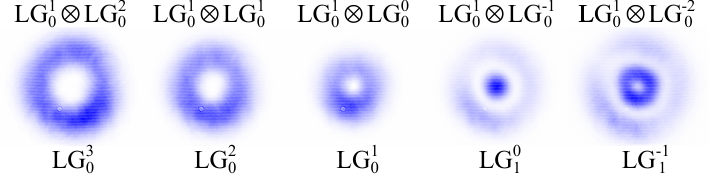}
\caption{Thick-medium blue FWM light arising from mixing $780\,$nm LG$_0^1$ with $776\,$nm LG$_0^{2,1,0,-1,-2}$ pump modes (left to right), with a transition from $p=0$ to $p=1$ when $\ell_{776}$ becomes negative (cf.\ Fig.~\ref{fig2:adder}).}
\label{fig:2b}
\end{figure}

For $p=0$ pump modes with the same intensity profile, but equal OAM of $\ell=1$, OAM conservation and Gouy phase matching restrict the generated blue light to an almost pure $\textrm{LG}_0^2$ mode, regardless of medium thickness (Supplementary Fig.~S2). In this situation, as well as similar experiments with pump modes of equal handed OAM, no conversion between angular and radial mode index occurs.
Investigating the onset of angular to radial mode conversion, we find that for $p=0$ pump beams, the blue light features a radial mode index of $p_\textrm{B}=(|\ell_{780}|+|\ell_{776}|-|\ell_{780}+\ell_{776}|)/2$, when working in a thick medium where Gouy phase matching applies (and for $u_\textrm{IR}=\textrm{LG}_0^0$). We demonstrate this for input modes $u_{780} \otimes u_{776} = \textrm{LG}^1_0 \otimes \textrm{LG}_0^{\ell_{776}}$, with $\ell_{776}\in\{-2,-1,0,1,2\}$, which transform cleanly into a blue LG mode, where $\ell_\textrm{B}=1+\ell_{776}$ and $p_\textrm{B}=0$ for $\ell_{776}\geq 0,$ and $p_\textrm{B}=1$ for $\ell_{776}<0$ (Fig.~\ref{fig:2b} and  Supplementary Fig.~S3).

\begin{figure}[!b]
\centering
\includegraphics[width=\columnwidth]{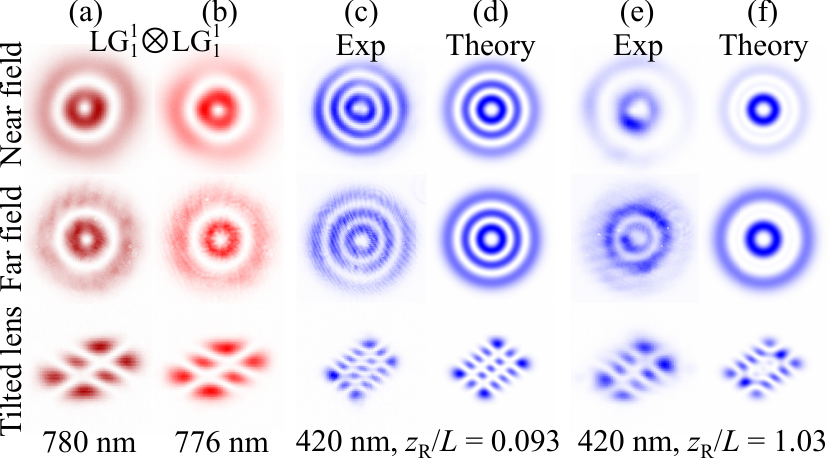}
\caption{FWM for both $780\,$nm pump (a) and $776\,$nm pump (b) in LG$_1^1$, with image triplets as defined in Fig.~\ref{fig2:adder}.  For a thick medium we observe the blue light in an almost pure LG$_2^2$ mode (c) as predicted (d).  For a thin medium however, the beam profile changes between near and far field (e), showing less agreement with our simple model (f).}
\label{fig3}
\end{figure}

In the following we extend our investigation to the mode conversion of pump beams with $p \neq 0$.  In Fig.~\ref{fig3} we show that pumping with $u_{780}=u_{776}=\textrm{LG}_1^1$ generates a clean LG$_2^2$ blue output mode if operating in the thick medium regime.  For a thin medium, a coherent superposition of modes with different mode orders is generated, leading to a changing intensity profile between near and far field. Here our simple theoretical model shows some qualitative agreement, but with notable discrepancies especially in the tilted lens decomposition. This may indicate that $u_\textrm{IR}$ is no longer constrained to the fundamental Gaussian mode, or that the Boyd criterion is a poor approximation in this case, as discussed in Supplementary Section S1.   

\begin{figure*}[!t]
\centering
\includegraphics[width=\linewidth]{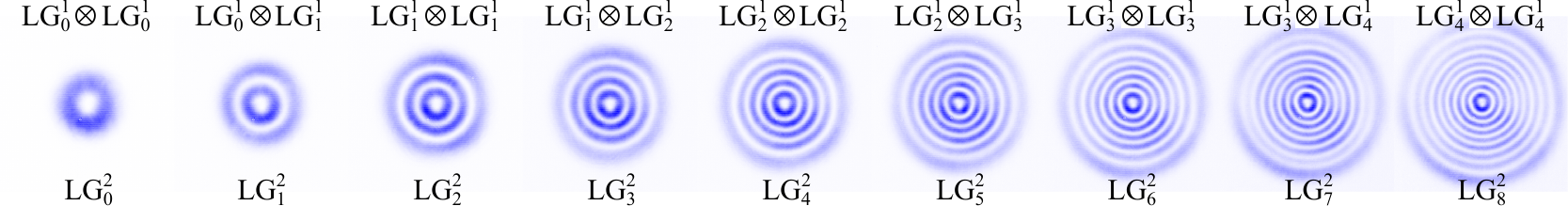}
\caption{Generation of higher order $p$ modes using FWM in the thick medium regime ($z_\textrm{R}/L=0.093$). Using a variety of $\mathrm{LG}_{p_{780}}^1 \otimes\mathrm{LG}_{p_{776}}^1$ pump mode combinations at $780\,$nm and  $776\,$nm, respectively, we generate blue light in modes with $\ell_\mathrm{B}=2$ and $0\leq p_\mathrm{B} \leq 8$, shown as near-field peak-normalised profiles.}
\label{fig:5}
\end{figure*}  

In the remainder of this Letter we are working in the thick medium regime where mode conversion is governed by Gouy phase matching and OAM conversion, as dictated by the 3D mode overlap. In this regime we find excellent agreement with our theoretical model when restricting the unobserved IR beam to $u_\textrm{IR}=\textrm{LG}_0^0$.
We demonstrate in Fig.~\ref{fig:5} the generation of blue light in LG$_p^2$, where $p$ takes all values from 0 to 8. We achieve this by using pump modes (Supplementary Fig.~S4) $u_{780} \otimes u_{776} = \textrm{LG}^1_{p_{780}} \otimes \textrm{LG}^1_{p_{776}}$, with $p_\mathrm{B}=p_{780}+p_{776}$. 
The $p$-index can be identified from the observed intensity profiles as the number of nodal rings.  We support this qualitative analysis by showing the associated high visibility interferograms (Supplementary Fig.~S5), formed by superposing the blue output beam with its mirror image, allowing us to determine $\ell$ by azimuthal Fourier decomposition.  Even at high $p$ values, the interferograms confirm that $\ell_\textrm{B}=2$, indicating high fidelity pump OAM transfer, without increasing the spiral bandwidth, i.e.\ without spreading the OAM distribution.

In conclusion, we have demonstrated a wide range of mode transformations facilitated by FWM in an atomic vapour, exploring the interplay between radial and azimuthal mode indices. Specifically, we have seen that input beams with $p=0$ and opposite $\ell$ values can generate a radial $p=1$ mode, while FWM light in higher order radial modes result from `adding' the $p$-values of the input modes. 
The generated mode decomposition critically depends on the medium thickness.  Gouy phase matching plays an increasing role as the length of the medium approaches and exceeds the Rayleigh range, thus limiting the number of participating eigenmodes. Our experiments are consistent, especially in the thick medium regime, with the assumption that the 5.2$\,\upmu$m light is generated only in the LG$^0_0$ mode.  Interestingly, this cannot be explained by the mode overlap integrals alone, and a conclusive explanation is still outstanding. 

We expect that our results are of relevance for nonlinear processes beyond FWM. In the thick-medium limit, new avenues of research exist both in `macroscopic' nonlinear optics (with crystals and atomic vapours), but also in nonlinear photonic waveguides \cite{Wang2019,Kim2020}, which can support LG beam modes  \cite{Chen20188,Chen2020} and where the extreme nonlinearities could lead to highly efficient on-chip wavelength and mode transformations. Most LG communication applications to date have been restricted to the azimuthal mode index, limiting the available information density. The possibility to convert efficiently between radial and azimuthal modes allows access to the full state-space for spatial mode encoding, with potential applications in all-optical communication and computing.

In the thin-medium limit \cite{Adams2020} we foresee that the addition of high $p$ modes may lead to high `radial bandwidth' entangled photons in much the same way that high $\ell$ leads to high spiral bandwidth \cite{Offer2018,Ding2019}, however correlation strength may depend on the degree of Gouy phase matching.   This is interesting both from the perspective of fundamental quantum concepts, as well as due to the increase of the potential state space available for application in optical quantum processes. 

There have been exciting recent results on extending resonant atomic four-wave mixing processes, as described in this paper, into the realms of deep-UV, mid-infrared and THz waves \cite{Antypas2019,Lam2019,Lam2020}. Clean mode conversion from the optical into these relatively inaccessible spectral regions, where reconfigurable optical elements like SLMs are not yet readily available, opens new research and development prospects for e.g.\ THz imaging \cite{Downes2020}.   

The dataset for this paper is available via Ref.~\cite{dataset}.\\  

\begin{acknowledgments}
We are grateful for valuable discussions with 
Irina Novikova, as well as for funding via the Leverhulme Trust (RPG-2013-386) and EPSRC (EP/M506643/1).
\end{acknowledgments}


\begin{thebibliography}{67}%
\makeatletter
\providecommand \@ifxundefined [1]{%
 \@ifx{#1\undefined}
}%
\providecommand \@ifnum [1]{%
 \ifnum #1\expandafter \@firstoftwo
 \else \expandafter \@secondoftwo
 \fi
}%
\providecommand \@ifx [1]{%
 \ifx #1\expandafter \@firstoftwo
 \else \expandafter \@secondoftwo
 \fi
}%
\providecommand \natexlab [1]{#1}%
\providecommand \enquote  [1]{``#1''}%
\providecommand \bibnamefont  [1]{#1}%
\providecommand \bibfnamefont [1]{#1}%
\providecommand \citenamefont [1]{#1}%
\providecommand \href@noop [0]{\@secondoftwo}%
\providecommand \href [0]{\begingroup \@sanitize@url \@href}%
\providecommand \@href[1]{\@@startlink{#1}\@@href}%
\providecommand \@@href[1]{\endgroup#1\@@endlink}%
\providecommand \@sanitize@url [0]{\catcode `\\12\catcode `\$12\catcode
  `\&12\catcode `\#12\catcode `\^12\catcode `\_12\catcode `\%12\relax}%
\providecommand \@@startlink[1]{}%
\providecommand \@@endlink[0]{}%
\providecommand \url  [0]{\begingroup\@sanitize@url \@url }%
\providecommand \@url [1]{\endgroup\@href {#1}{\urlprefix }}%
\providecommand \urlprefix  [0]{URL }%
\providecommand \Eprint [0]{\href }%
\providecommand \doibase [0]{https://doi.org/}%
\providecommand \selectlanguage [0]{\@gobble}%
\providecommand \bibinfo  [0]{\@secondoftwo}%
\providecommand \bibfield  [0]{\@secondoftwo}%
\providecommand \translation [1]{[#1]}%
\providecommand \BibitemOpen [0]{}%
\providecommand \bibitemStop [0]{}%
\providecommand \bibitemNoStop [0]{.\EOS\space}%
\providecommand \EOS [0]{\spacefactor3000\relax}%
\providecommand \BibitemShut  [1]{\csname bibitem#1\endcsname}%
\let\auto@bib@innerbib\@empty
\bibitem [{\citenamefont {Allen}\ \emph {et~al.}(1992)\citenamefont {Allen},
  \citenamefont {Beijersbergen}, \citenamefont {Spreeuw},\ and\ \citenamefont
  {Woerdman}}]{Allen1992}%
  \BibitemOpen
  \bibfield  {author} {\bibinfo {author} {\bibfnamefont {L.}~\bibnamefont
  {Allen}}, \bibinfo {author} {\bibfnamefont {M.~W.}\ \bibnamefont
  {Beijersbergen}}, \bibinfo {author} {\bibfnamefont {R.~J.~C.}\ \bibnamefont
  {Spreeuw}},\ and\ \bibinfo {author} {\bibfnamefont {J.~P.}\ \bibnamefont
  {Woerdman}},\ }\href {https://doi.org/10.1103/PhysRevA.45.8185} {\bibfield
  {journal} {\bibinfo  {journal} {Phys. Rev. A}\ }\textbf {\bibinfo {volume}
  {45}},\ \bibinfo {pages} {8185} (\bibinfo {year} {1992})}\BibitemShut
  {NoStop}%
\bibitem [{\citenamefont {Ni}\ \emph {et~al.}(2017)\citenamefont {Ni},
  \citenamefont {Wang}, \citenamefont {Li}, \citenamefont {Lao}, \citenamefont
  {Hu}, \citenamefont {Ji}, \citenamefont {Xu}, \citenamefont {Zhang},
  \citenamefont {Li}, \citenamefont {Wu},\ and\ \citenamefont {Chu}}]{Ni2017}%
  \BibitemOpen
  \bibfield  {author} {\bibinfo {author} {\bibfnamefont {J.}~\bibnamefont
  {Ni}}, \bibinfo {author} {\bibfnamefont {Z.}~\bibnamefont {Wang}}, \bibinfo
  {author} {\bibfnamefont {Z.}~\bibnamefont {Li}}, \bibinfo {author}
  {\bibfnamefont {Z.}~\bibnamefont {Lao}}, \bibinfo {author} {\bibfnamefont
  {Y.}~\bibnamefont {Hu}}, \bibinfo {author} {\bibfnamefont {S.}~\bibnamefont
  {Ji}}, \bibinfo {author} {\bibfnamefont {B.}~\bibnamefont {Xu}}, \bibinfo
  {author} {\bibfnamefont {C.}~\bibnamefont {Zhang}}, \bibinfo {author}
  {\bibfnamefont {J.}~\bibnamefont {Li}}, \bibinfo {author} {\bibfnamefont
  {D.}~\bibnamefont {Wu}},\ and\ \bibinfo {author} {\bibfnamefont
  {J.}~\bibnamefont {Chu}},\ }\href {https://doi.org/10.1002/adfm.201701939}
  {\bibfield  {journal} {\bibinfo  {journal} {Adv.\ Funct.\ Mater.}\ }\textbf
  {\bibinfo {volume} {27}},\ \bibinfo {pages} {1701939} (\bibinfo {year}
  {2017})}\BibitemShut {NoStop}%
\bibitem [{\citenamefont {Pan}\ \emph {et~al.}(2019)\citenamefont {Pan},
  \citenamefont {Yu}, \citenamefont {Zhou}, \citenamefont {Zhang},
  \citenamefont {Zhang}, \citenamefont {Lv}, \citenamefont {Li}, \citenamefont
  {Wang},\ and\ \citenamefont {Jing}}]{Pan2019}%
  \BibitemOpen
  \bibfield  {author} {\bibinfo {author} {\bibfnamefont {X.}~\bibnamefont
  {Pan}}, \bibinfo {author} {\bibfnamefont {S.}~\bibnamefont {Yu}}, \bibinfo
  {author} {\bibfnamefont {Y.}~\bibnamefont {Zhou}}, \bibinfo {author}
  {\bibfnamefont {K.}~\bibnamefont {Zhang}}, \bibinfo {author} {\bibfnamefont
  {K.}~\bibnamefont {Zhang}}, \bibinfo {author} {\bibfnamefont
  {S.}~\bibnamefont {Lv}}, \bibinfo {author} {\bibfnamefont {S.}~\bibnamefont
  {Li}}, \bibinfo {author} {\bibfnamefont {W.}~\bibnamefont {Wang}},\ and\
  \bibinfo {author} {\bibfnamefont {J.}~\bibnamefont {Jing}},\ }\href
  {https://doi.org/10.1103/physrevlett.123.070506} {\bibfield  {journal}
  {\bibinfo  {journal} {Phys.\ Rev.\ Lett.}\ }\textbf {\bibinfo {volume}
  {123}},\ \bibinfo {pages} {070506} (\bibinfo {year} {2019})}\BibitemShut
  {NoStop}%
\bibitem [{\citenamefont {Ding}\ \emph {et~al.}(2016)\citenamefont {Ding},
  \citenamefont {Zhang}, \citenamefont {Shi}, \citenamefont {Zhou},
  \citenamefont {Li}, \citenamefont {Shi},\ and\ \citenamefont
  {Guo}}]{Ding2016}%
  \BibitemOpen
  \bibfield  {author} {\bibinfo {author} {\bibfnamefont {D.-S.}\ \bibnamefont
  {Ding}}, \bibinfo {author} {\bibfnamefont {W.}~\bibnamefont {Zhang}},
  \bibinfo {author} {\bibfnamefont {S.}~\bibnamefont {Shi}}, \bibinfo {author}
  {\bibfnamefont {Z.-Y.}\ \bibnamefont {Zhou}}, \bibinfo {author}
  {\bibfnamefont {Y.}~\bibnamefont {Li}}, \bibinfo {author} {\bibfnamefont
  {B.-S.}\ \bibnamefont {Shi}},\ and\ \bibinfo {author} {\bibfnamefont {G.-C.}\
  \bibnamefont {Guo}},\ }\href {https://doi.org/10.1038/lsa.2016.157}
  {\bibfield  {journal} {\bibinfo  {journal} {Light Sci.\ Appl.}\ }\textbf
  {\bibinfo {volume} {5}},\ \bibinfo {pages} {e16157} (\bibinfo {year}
  {2016})}\BibitemShut {NoStop}%
\bibitem [{\citenamefont {Gibson}\ \emph {et~al.}(2004)\citenamefont {Gibson},
  \citenamefont {Courtial}, \citenamefont {Padgett}, \citenamefont {Vasnetsov},
  \citenamefont {Pas'ko}, \citenamefont {Barnett},\ and\ \citenamefont
  {Franke-Arnold}}]{Gibson2004}%
  \BibitemOpen
  \bibfield  {author} {\bibinfo {author} {\bibfnamefont {G.}~\bibnamefont
  {Gibson}}, \bibinfo {author} {\bibfnamefont {J.}~\bibnamefont {Courtial}},
  \bibinfo {author} {\bibfnamefont {M.~J.}\ \bibnamefont {Padgett}}, \bibinfo
  {author} {\bibfnamefont {M.}~\bibnamefont {Vasnetsov}}, \bibinfo {author}
  {\bibfnamefont {V.}~\bibnamefont {Pas'ko}}, \bibinfo {author} {\bibfnamefont
  {S.~M.}\ \bibnamefont {Barnett}},\ and\ \bibinfo {author} {\bibfnamefont
  {S.}~\bibnamefont {Franke-Arnold}},\ }\href
  {https://doi.org/10.1364/opex.12.005448} {\bibfield  {journal} {\bibinfo
  {journal} {Opt.\ Express}\ }\textbf {\bibinfo {volume} {12}},\ \bibinfo
  {pages} {5448} (\bibinfo {year} {2004})}\BibitemShut {NoStop}%
\bibitem [{\citenamefont {Yue}\ \emph {et~al.}(2018)\citenamefont {Yue},
  \citenamefont {Ren}, \citenamefont {Wei}, \citenamefont {Lin},\ and\
  \citenamefont {Gu}}]{Yue2018}%
  \BibitemOpen
  \bibfield  {author} {\bibinfo {author} {\bibfnamefont {Z.}~\bibnamefont
  {Yue}}, \bibinfo {author} {\bibfnamefont {H.}~\bibnamefont {Ren}}, \bibinfo
  {author} {\bibfnamefont {S.}~\bibnamefont {Wei}}, \bibinfo {author}
  {\bibfnamefont {J.}~\bibnamefont {Lin}},\ and\ \bibinfo {author}
  {\bibfnamefont {M.}~\bibnamefont {Gu}},\ }\href
  {https://doi.org/10.1038/s41467-018-06952-1} {\bibfield  {journal} {\bibinfo
  {journal} {Nat.\ Commun.}\ }\textbf {\bibinfo {volume} {9}},\ \bibinfo
  {pages} {4413} (\bibinfo {year} {2018})}\BibitemShut {NoStop}%
\bibitem [{\citenamefont {Bozinovic}\ \emph {et~al.}(2013)\citenamefont
  {Bozinovic}, \citenamefont {Yue}, \citenamefont {Ren}, \citenamefont {Tur},
  \citenamefont {Kristensen}, \citenamefont {Huang}, \citenamefont {Willner},\
  and\ \citenamefont {Ramachandran}}]{Bozinovic2013}%
  \BibitemOpen
  \bibfield  {author} {\bibinfo {author} {\bibfnamefont {N.}~\bibnamefont
  {Bozinovic}}, \bibinfo {author} {\bibfnamefont {Y.}~\bibnamefont {Yue}},
  \bibinfo {author} {\bibfnamefont {Y.}~\bibnamefont {Ren}}, \bibinfo {author}
  {\bibfnamefont {M.}~\bibnamefont {Tur}}, \bibinfo {author} {\bibfnamefont
  {P.}~\bibnamefont {Kristensen}}, \bibinfo {author} {\bibfnamefont
  {H.}~\bibnamefont {Huang}}, \bibinfo {author} {\bibfnamefont {A.~E.}\
  \bibnamefont {Willner}},\ and\ \bibinfo {author} {\bibfnamefont
  {S.}~\bibnamefont {Ramachandran}},\ }\href
  {https://doi.org/10.1126/science.1237861} {\bibfield  {journal} {\bibinfo
  {journal} {Science}\ }\textbf {\bibinfo {volume} {340}},\ \bibinfo {pages}
  {1545} (\bibinfo {year} {2013})}\BibitemShut {NoStop}%
\bibitem [{\citenamefont {Yao}\ and\ \citenamefont
  {Padgett}(2011)}]{Yao2011rev}%
  \BibitemOpen
  \bibfield  {author} {\bibinfo {author} {\bibfnamefont {A.~M.}\ \bibnamefont
  {Yao}}\ and\ \bibinfo {author} {\bibfnamefont {M.~J.}\ \bibnamefont
  {Padgett}},\ }\href {https://doi.org/10.1364/aop.3.000161} {\bibfield
  {journal} {\bibinfo  {journal} {Adv.\ Opt.\ Phot.}\ }\textbf {\bibinfo
  {volume} {3}},\ \bibinfo {pages} {161} (\bibinfo {year} {2011})}\BibitemShut
  {NoStop}%
\bibitem [{\citenamefont {Bliokh}\ \emph {et~al.}(2015)\citenamefont {Bliokh},
  \citenamefont {Rodr{\'{\i}}guez-Fortu{\~{n}}o}, \citenamefont {Nori},\ and\
  \citenamefont {Zayats}}]{Bliokh2015}%
  \BibitemOpen
  \bibfield  {author} {\bibinfo {author} {\bibfnamefont {K.~Y.}\ \bibnamefont
  {Bliokh}}, \bibinfo {author} {\bibfnamefont {F.~J.}\ \bibnamefont
  {Rodr{\'{\i}}guez-Fortu{\~{n}}o}}, \bibinfo {author} {\bibfnamefont
  {F.}~\bibnamefont {Nori}},\ and\ \bibinfo {author} {\bibfnamefont {A.~V.}\
  \bibnamefont {Zayats}},\ }\href {https://doi.org/10.1038/nphoton.2015.201}
  {\bibfield  {journal} {\bibinfo  {journal} {Nat.\ Photonics}\ }\textbf
  {\bibinfo {volume} {9}},\ \bibinfo {pages} {796} (\bibinfo {year}
  {2015})}\BibitemShut {NoStop}%
\bibitem [{\citenamefont {Rubinsztein-Dunlop}\ \emph
  {et~al.}(2016)\citenamefont {Rubinsztein-Dunlop}, \citenamefont {Forbes},
  \citenamefont {Berry}, \citenamefont {Dennis}, \citenamefont {Andrews},
  \citenamefont {Mansuripur}, \citenamefont {Denz}, \citenamefont {Alpmann},
  \citenamefont {Banzer}, \citenamefont {Bauer}, \citenamefont {Karimi},
  \citenamefont {Marrucci}, \citenamefont {Padgett}, \citenamefont
  {Ritsch-Marte}, \citenamefont {Litchinitser}, \citenamefont {Bigelow},
  \citenamefont {Rosales-Guzm{\'{a}}n}, \citenamefont {Belmonte}, \citenamefont
  {Torres}, \citenamefont {Neely}, \citenamefont {Baker}, \citenamefont
  {Gordon}, \citenamefont {Stilgoe}, \citenamefont {Romero}, \citenamefont
  {White}, \citenamefont {Fickler}, \citenamefont {Willner}, \citenamefont
  {Xie}, \citenamefont {McMorran},\ and\ \citenamefont
  {Weiner}}]{Rubinsztein2017}%
  \BibitemOpen
  \bibfield  {author} {\bibinfo {author} {\bibfnamefont {H.}~\bibnamefont
  {Rubinsztein-Dunlop}}, \bibinfo {author} {\bibfnamefont {A.}~\bibnamefont
  {Forbes}}, \bibinfo {author} {\bibfnamefont {M.~V.}\ \bibnamefont {Berry}},
  \bibinfo {author} {\bibfnamefont {M.~R.}\ \bibnamefont {Dennis}}, \bibinfo
  {author} {\bibfnamefont {D.~L.}\ \bibnamefont {Andrews}}, \bibinfo {author}
  {\bibfnamefont {M.}~\bibnamefont {Mansuripur}}, \bibinfo {author}
  {\bibfnamefont {C.}~\bibnamefont {Denz}}, \bibinfo {author} {\bibfnamefont
  {C.}~\bibnamefont {Alpmann}}, \bibinfo {author} {\bibfnamefont
  {P.}~\bibnamefont {Banzer}}, \bibinfo {author} {\bibfnamefont
  {T.}~\bibnamefont {Bauer}}, \bibinfo {author} {\bibfnamefont
  {E.}~\bibnamefont {Karimi}}, \bibinfo {author} {\bibfnamefont
  {L.}~\bibnamefont {Marrucci}}, \bibinfo {author} {\bibfnamefont
  {M.}~\bibnamefont {Padgett}}, \bibinfo {author} {\bibfnamefont
  {M.}~\bibnamefont {Ritsch-Marte}}, \bibinfo {author} {\bibfnamefont {N.~M.}\
  \bibnamefont {Litchinitser}}, \bibinfo {author} {\bibfnamefont {N.~P.}\
  \bibnamefont {Bigelow}}, \bibinfo {author} {\bibfnamefont {C.}~\bibnamefont
  {Rosales-Guzm{\'{a}}n}}, \bibinfo {author} {\bibfnamefont {A.}~\bibnamefont
  {Belmonte}}, \bibinfo {author} {\bibfnamefont {J.~P.}\ \bibnamefont
  {Torres}}, \bibinfo {author} {\bibfnamefont {T.~W.}\ \bibnamefont {Neely}},
  \bibinfo {author} {\bibfnamefont {M.}~\bibnamefont {Baker}}, \bibinfo
  {author} {\bibfnamefont {R.}~\bibnamefont {Gordon}}, \bibinfo {author}
  {\bibfnamefont {A.~B.}\ \bibnamefont {Stilgoe}}, \bibinfo {author}
  {\bibfnamefont {J.}~\bibnamefont {Romero}}, \bibinfo {author} {\bibfnamefont
  {A.~G.}\ \bibnamefont {White}}, \bibinfo {author} {\bibfnamefont
  {R.}~\bibnamefont {Fickler}}, \bibinfo {author} {\bibfnamefont {A.~E.}\
  \bibnamefont {Willner}}, \bibinfo {author} {\bibfnamefont {G.}~\bibnamefont
  {Xie}}, \bibinfo {author} {\bibfnamefont {B.}~\bibnamefont {McMorran}},\ and\
  \bibinfo {author} {\bibfnamefont {A.~M.}\ \bibnamefont {Weiner}},\ }\href
  {https://doi.org/10.1088/2040-8978/19/1/013001} {\bibfield  {journal}
  {\bibinfo  {journal} {J.\ Opt.}\ }\textbf {\bibinfo {volume} {19}},\ \bibinfo
  {pages} {013001} (\bibinfo {year} {2016})}\BibitemShut {NoStop}%
\bibitem [{\citenamefont {Franke-Arnold}(2017)}]{FrankeArnold2017}%
  \BibitemOpen
  \bibfield  {author} {\bibinfo {author} {\bibfnamefont {S.}~\bibnamefont
  {Franke-Arnold}},\ }\href {https://doi.org/10.1098/rsta.2015.0435} {\bibfield
   {journal} {\bibinfo  {journal} {Phil.\ Trans.\ Royal Soc.\ A}\ }\textbf
  {\bibinfo {volume} {375}},\ \bibinfo {pages} {20150435} (\bibinfo {year}
  {2017})}\BibitemShut {NoStop}%
\bibitem [{\citenamefont {Karimi}\ and\ \citenamefont
  {Santamato}(2012)}]{Karimi2012}%
  \BibitemOpen
  \bibfield  {author} {\bibinfo {author} {\bibfnamefont {E.}~\bibnamefont
  {Karimi}}\ and\ \bibinfo {author} {\bibfnamefont {E.}~\bibnamefont
  {Santamato}},\ }\href {https://doi.org/10.1364/ol.37.002484} {\bibfield
  {journal} {\bibinfo  {journal} {Optics Letters}\ }\textbf {\bibinfo {volume}
  {37}},\ \bibinfo {pages} {2484} (\bibinfo {year} {2012})}\BibitemShut
  {NoStop}%
\bibitem [{\citenamefont {Mazilu}\ \emph {et~al.}(2012)\citenamefont {Mazilu},
  \citenamefont {Mourka}, \citenamefont {Vettenburg}, \citenamefont {Wright},\
  and\ \citenamefont {Dholakia}}]{Dholakia2012}%
  \BibitemOpen
  \bibfield  {author} {\bibinfo {author} {\bibfnamefont {M.}~\bibnamefont
  {Mazilu}}, \bibinfo {author} {\bibfnamefont {A.}~\bibnamefont {Mourka}},
  \bibinfo {author} {\bibfnamefont {T.}~\bibnamefont {Vettenburg}}, \bibinfo
  {author} {\bibfnamefont {E.~M.}\ \bibnamefont {Wright}},\ and\ \bibinfo
  {author} {\bibfnamefont {K.}~\bibnamefont {Dholakia}},\ }\href
  {https://doi.org/10.1063/1.4728111} {\bibfield  {journal} {\bibinfo
  {journal} {Appl.\ Phys.\ Lett.}\ }\textbf {\bibinfo {volume} {100}},\
  \bibinfo {pages} {231115} (\bibinfo {year} {2012})}\BibitemShut {NoStop}%
\bibitem [{\citenamefont {Dudley}\ \emph {et~al.}(2013)\citenamefont {Dudley},
  \citenamefont {Mhlanga}, \citenamefont {Lavery}, \citenamefont {McDonald},
  \citenamefont {Roux}, \citenamefont {Padgett},\ and\ \citenamefont
  {Forbes}}]{Dudley2013}%
  \BibitemOpen
  \bibfield  {author} {\bibinfo {author} {\bibfnamefont {A.}~\bibnamefont
  {Dudley}}, \bibinfo {author} {\bibfnamefont {T.}~\bibnamefont {Mhlanga}},
  \bibinfo {author} {\bibfnamefont {M.}~\bibnamefont {Lavery}}, \bibinfo
  {author} {\bibfnamefont {A.}~\bibnamefont {McDonald}}, \bibinfo {author}
  {\bibfnamefont {F.~S.}\ \bibnamefont {Roux}}, \bibinfo {author}
  {\bibfnamefont {M.}~\bibnamefont {Padgett}},\ and\ \bibinfo {author}
  {\bibfnamefont {A.}~\bibnamefont {Forbes}},\ }\href
  {https://doi.org/10.1364/OE.21.000165} {\bibfield  {journal} {\bibinfo
  {journal} {Opt. Express}\ }\textbf {\bibinfo {volume} {21}},\ \bibinfo
  {pages} {165} (\bibinfo {year} {2013})}\BibitemShut {NoStop}%
\bibitem [{\citenamefont {Zhou}\ \emph {et~al.}(2016)\citenamefont {Zhou},
  \citenamefont {Zhang},\ and\ \citenamefont {Chen}}]{Zhou2016APL}%
  \BibitemOpen
  \bibfield  {author} {\bibinfo {author} {\bibfnamefont {J.}~\bibnamefont
  {Zhou}}, \bibinfo {author} {\bibfnamefont {W.}~\bibnamefont {Zhang}},\ and\
  \bibinfo {author} {\bibfnamefont {L.}~\bibnamefont {Chen}},\ }\href
  {https://doi.org/10.1063/1.4944463} {\bibfield  {journal} {\bibinfo
  {journal} {Appl.\ Phys.\ Lett.}\ }\textbf {\bibinfo {volume} {108}},\
  \bibinfo {pages} {111108} (\bibinfo {year} {2016})}\BibitemShut {NoStop}%
\bibitem [{\citenamefont {Fontaine}\ \emph {et~al.}(2019)\citenamefont
  {Fontaine}, \citenamefont {Ryf}, \citenamefont {Chen}, \citenamefont
  {Neilson}, \citenamefont {Kim},\ and\ \citenamefont
  {Carpenter}}]{Fontaine2019}%
  \BibitemOpen
  \bibfield  {author} {\bibinfo {author} {\bibfnamefont {N.~K.}\ \bibnamefont
  {Fontaine}}, \bibinfo {author} {\bibfnamefont {R.}~\bibnamefont {Ryf}},
  \bibinfo {author} {\bibfnamefont {H.}~\bibnamefont {Chen}}, \bibinfo {author}
  {\bibfnamefont {D.~T.}\ \bibnamefont {Neilson}}, \bibinfo {author}
  {\bibfnamefont {K.}~\bibnamefont {Kim}},\ and\ \bibinfo {author}
  {\bibfnamefont {J.}~\bibnamefont {Carpenter}},\ }\href
  {https://doi.org/10.1038/s41467-019-09840-4} {\bibfield  {journal} {\bibinfo
  {journal} {Nat.\ Commun.}\ }\textbf {\bibinfo {volume} {10}},\ \bibinfo
  {pages} {1865} (\bibinfo {year} {2019})}\BibitemShut {NoStop}%
\bibitem [{\citenamefont {Zhou}\ \emph {et~al.}(2017)\citenamefont {Zhou},
  \citenamefont {Mirhosseini}, \citenamefont {Fu}, \citenamefont {Zhao},
  \citenamefont {Rafsanjani}, \citenamefont {Willner},\ and\ \citenamefont
  {Boyd}}]{Boyd2017}%
  \BibitemOpen
  \bibfield  {author} {\bibinfo {author} {\bibfnamefont {Y.}~\bibnamefont
  {Zhou}}, \bibinfo {author} {\bibfnamefont {M.}~\bibnamefont {Mirhosseini}},
  \bibinfo {author} {\bibfnamefont {D.}~\bibnamefont {Fu}}, \bibinfo {author}
  {\bibfnamefont {J.}~\bibnamefont {Zhao}}, \bibinfo {author} {\bibfnamefont
  {S.~M.~H.}\ \bibnamefont {Rafsanjani}}, \bibinfo {author} {\bibfnamefont
  {A.~E.}\ \bibnamefont {Willner}},\ and\ \bibinfo {author} {\bibfnamefont
  {R.~W.}\ \bibnamefont {Boyd}},\ }\href
  {https://doi.org/10.1103/physrevlett.119.263602} {\bibfield  {journal}
  {\bibinfo  {journal} {Phys.\ Rev.\ Lett.}\ }\textbf {\bibinfo {volume}
  {119}},\ \bibinfo {pages} {263602} (\bibinfo {year} {2017})}\BibitemShut
  {NoStop}%
\bibitem [{\citenamefont {Gu}\ \emph {et~al.}(2018)\citenamefont {Gu},
  \citenamefont {Krenn}, \citenamefont {Erhard},\ and\ \citenamefont
  {Zeilinger}}]{Zeilinger2018}%
  \BibitemOpen
  \bibfield  {author} {\bibinfo {author} {\bibfnamefont {X.}~\bibnamefont
  {Gu}}, \bibinfo {author} {\bibfnamefont {M.}~\bibnamefont {Krenn}}, \bibinfo
  {author} {\bibfnamefont {M.}~\bibnamefont {Erhard}},\ and\ \bibinfo {author}
  {\bibfnamefont {A.}~\bibnamefont {Zeilinger}},\ }\href
  {https://doi.org/10.1103/physrevlett.120.103601} {\bibfield  {journal}
  {\bibinfo  {journal} {Phys.\ Rev.\ Lett.}\ }\textbf {\bibinfo {volume}
  {120}},\ \bibinfo {pages} {103601} (\bibinfo {year} {2018})}\BibitemShut
  {NoStop}%
\bibitem [{\citenamefont {Salakhutdinov}\ \emph {et~al.}(2012)\citenamefont
  {Salakhutdinov}, \citenamefont {Eliel},\ and\ \citenamefont
  {L\"{o}ffler}}]{Loeffler2012}%
  \BibitemOpen
  \bibfield  {author} {\bibinfo {author} {\bibfnamefont {V.~D.}\ \bibnamefont
  {Salakhutdinov}}, \bibinfo {author} {\bibfnamefont {E.~R.}\ \bibnamefont
  {Eliel}},\ and\ \bibinfo {author} {\bibfnamefont {W.}~\bibnamefont
  {L\"{o}ffler}},\ }\href {https://doi.org/10.1103/physrevlett.108.173604}
  {\bibfield  {journal} {\bibinfo  {journal} {Phys.\ Rev.\ Lett.}\ }\textbf
  {\bibinfo {volume} {108}},\ \bibinfo {pages} {173604} (\bibinfo {year}
  {2012})}\BibitemShut {NoStop}%
\bibitem [{\citenamefont {Karimi}\ \emph
  {et~al.}(2014{\natexlab{a}})\citenamefont {Karimi}, \citenamefont
  {Giovannini}, \citenamefont {Bolduc}, \citenamefont {Bent}, \citenamefont
  {Miatto}, \citenamefont {Padgett},\ and\ \citenamefont {Boyd}}]{Boyd2014HOM}%
  \BibitemOpen
  \bibfield  {author} {\bibinfo {author} {\bibfnamefont {E.}~\bibnamefont
  {Karimi}}, \bibinfo {author} {\bibfnamefont {D.}~\bibnamefont {Giovannini}},
  \bibinfo {author} {\bibfnamefont {E.}~\bibnamefont {Bolduc}}, \bibinfo
  {author} {\bibfnamefont {N.}~\bibnamefont {Bent}}, \bibinfo {author}
  {\bibfnamefont {F.~M.}\ \bibnamefont {Miatto}}, \bibinfo {author}
  {\bibfnamefont {M.~J.}\ \bibnamefont {Padgett}},\ and\ \bibinfo {author}
  {\bibfnamefont {R.~W.}\ \bibnamefont {Boyd}},\ }\href
  {https://doi.org/10.1103/physreva.89.013829} {\bibfield  {journal} {\bibinfo
  {journal} {Phys.\ Rev.\ A}\ }\textbf {\bibinfo {volume} {89}},\ \bibinfo
  {pages} {013829} (\bibinfo {year} {2014}{\natexlab{a}})}\BibitemShut
  {NoStop}%
\bibitem [{\citenamefont {Karimi}\ \emph
  {et~al.}(2014{\natexlab{b}})\citenamefont {Karimi}, \citenamefont {Boyd},
  \citenamefont {de~la Hoz}, \citenamefont {de~Guise}, \citenamefont
  {{\v{R}}eh{\'{a}}{\v{c}}ek}, \citenamefont {Hradil}, \citenamefont {Aiello},
  \citenamefont {Leuchs},\ and\ \citenamefont {S{\'{a}}nchez-Soto}}]{Boyd2014}%
  \BibitemOpen
  \bibfield  {author} {\bibinfo {author} {\bibfnamefont {E.}~\bibnamefont
  {Karimi}}, \bibinfo {author} {\bibfnamefont {R.~W.}\ \bibnamefont {Boyd}},
  \bibinfo {author} {\bibfnamefont {P.}~\bibnamefont {de~la Hoz}}, \bibinfo
  {author} {\bibfnamefont {H.}~\bibnamefont {de~Guise}}, \bibinfo {author}
  {\bibfnamefont {J.}~\bibnamefont {{\v{R}}eh{\'{a}}{\v{c}}ek}}, \bibinfo
  {author} {\bibfnamefont {Z.}~\bibnamefont {Hradil}}, \bibinfo {author}
  {\bibfnamefont {A.}~\bibnamefont {Aiello}}, \bibinfo {author} {\bibfnamefont
  {G.}~\bibnamefont {Leuchs}},\ and\ \bibinfo {author} {\bibfnamefont {L.~L.}\
  \bibnamefont {S{\'{a}}nchez-Soto}},\ }\href
  {https://doi.org/10.1103/physreva.89.063813} {\bibfield  {journal} {\bibinfo
  {journal} {Phys.\ Rev.\ A}\ }\textbf {\bibinfo {volume} {89}},\ \bibinfo
  {pages} {063813} (\bibinfo {year} {2014}{\natexlab{b}})}\BibitemShut
  {NoStop}%
\bibitem [{\citenamefont {Plick}\ and\ \citenamefont
  {Krenn}(2015)}]{Krenn2015}%
  \BibitemOpen
  \bibfield  {author} {\bibinfo {author} {\bibfnamefont {W.~N.}\ \bibnamefont
  {Plick}}\ and\ \bibinfo {author} {\bibfnamefont {M.}~\bibnamefont {Krenn}},\
  }\href {https://doi.org/10.1103/physreva.92.063841} {\bibfield  {journal}
  {\bibinfo  {journal} {Phys.\ Rev.\ A}\ }\textbf {\bibinfo {volume} {92}},\
  \bibinfo {pages} {063841} (\bibinfo {year} {2015})}\BibitemShut {NoStop}%
\bibitem [{\citenamefont {Zhang}\ \emph {et~al.}(2018)\citenamefont {Zhang},
  \citenamefont {Qiu}, \citenamefont {Zhang},\ and\ \citenamefont
  {Chen}}]{Zhang2018}%
  \BibitemOpen
  \bibfield  {author} {\bibinfo {author} {\bibfnamefont {D.}~\bibnamefont
  {Zhang}}, \bibinfo {author} {\bibfnamefont {X.}~\bibnamefont {Qiu}}, \bibinfo
  {author} {\bibfnamefont {W.}~\bibnamefont {Zhang}},\ and\ \bibinfo {author}
  {\bibfnamefont {L.}~\bibnamefont {Chen}},\ }\href
  {https://doi.org/10.1103/physreva.98.042134} {\bibfield  {journal} {\bibinfo
  {journal} {Phys.\ Rev.\ A}\ }\textbf {\bibinfo {volume} {98}},\ \bibinfo
  {pages} {042134} (\bibinfo {year} {2018})}\BibitemShut {NoStop}%
\bibitem [{\citenamefont {Chen}\ \emph {et~al.}(2019)\citenamefont {Chen},
  \citenamefont {Ma}, \citenamefont {Qiu}, \citenamefont {Zhang}, \citenamefont
  {Zhang},\ and\ \citenamefont {Boyd}}]{Boyd2019}%
  \BibitemOpen
  \bibfield  {author} {\bibinfo {author} {\bibfnamefont {L.}~\bibnamefont
  {Chen}}, \bibinfo {author} {\bibfnamefont {T.}~\bibnamefont {Ma}}, \bibinfo
  {author} {\bibfnamefont {X.}~\bibnamefont {Qiu}}, \bibinfo {author}
  {\bibfnamefont {D.}~\bibnamefont {Zhang}}, \bibinfo {author} {\bibfnamefont
  {W.}~\bibnamefont {Zhang}},\ and\ \bibinfo {author} {\bibfnamefont {R.~W.}\
  \bibnamefont {Boyd}},\ }\href
  {https://doi.org/10.1103/physrevlett.123.060403} {\bibfield  {journal}
  {\bibinfo  {journal} {Phys.\ Rev.\ Lett.}\ }\textbf {\bibinfo {volume}
  {123}},\ \bibinfo {pages} {060403} (\bibinfo {year} {2019})}\BibitemShut
  {NoStop}%
\bibitem [{\citenamefont {Yao}(2011)}]{Yao2011}%
  \BibitemOpen
  \bibfield  {author} {\bibinfo {author} {\bibfnamefont {A.~M.}\ \bibnamefont
  {Yao}},\ }\href {https://doi.org/10.1088/1367-2630/13/5/053048} {\bibfield
  {journal} {\bibinfo  {journal} {New J.\ Phys.}\ }\textbf {\bibinfo {volume}
  {13}},\ \bibinfo {pages} {053048} (\bibinfo {year} {2011})}\BibitemShut
  {NoStop}%
\bibitem [{\citenamefont {Shao}\ \emph {et~al.}(2013)\citenamefont {Shao},
  \citenamefont {Wu}, \citenamefont {Chen}, \citenamefont {Xu},\ and\
  \citenamefont {Lu}}]{Shao2013}%
  \BibitemOpen
  \bibfield  {author} {\bibinfo {author} {\bibfnamefont {G.~H.}\ \bibnamefont
  {Shao}}, \bibinfo {author} {\bibfnamefont {Z.~J.}\ \bibnamefont {Wu}},
  \bibinfo {author} {\bibfnamefont {J.~H.}\ \bibnamefont {Chen}}, \bibinfo
  {author} {\bibfnamefont {F.}~\bibnamefont {Xu}},\ and\ \bibinfo {author}
  {\bibfnamefont {Y.~Q.}\ \bibnamefont {Lu}},\ }\href
  {https://doi.org/10.1103/PhysRevA.88.063827} {\bibfield  {journal} {\bibinfo
  {journal} {Phys.\ Rev.\ A}\ }\textbf {\bibinfo {volume} {88}},\ \bibinfo
  {pages} {063827} (\bibinfo {year} {2013})}\BibitemShut {NoStop}%
\bibitem [{\citenamefont {Courtial}\ \emph {et~al.}(1997)\citenamefont
  {Courtial}, \citenamefont {Dholakia}, \citenamefont {Allen},\ and\
  \citenamefont {Padgett}}]{Courtial1997}%
  \BibitemOpen
  \bibfield  {author} {\bibinfo {author} {\bibfnamefont {J.}~\bibnamefont
  {Courtial}}, \bibinfo {author} {\bibfnamefont {K.}~\bibnamefont {Dholakia}},
  \bibinfo {author} {\bibfnamefont {L.}~\bibnamefont {Allen}},\ and\ \bibinfo
  {author} {\bibfnamefont {M.~J.}\ \bibnamefont {Padgett}},\ }\href
  {https://doi.org/10.1103/PhysRevA.56.4193} {\bibfield  {journal} {\bibinfo
  {journal} {Phys. Rev. A}\ }\textbf {\bibinfo {volume} {56}},\ \bibinfo
  {pages} {4193} (\bibinfo {year} {1997})}\BibitemShut {NoStop}%
\bibitem [{\citenamefont {Mair}\ \emph {et~al.}(2001)\citenamefont {Mair},
  \citenamefont {Vaziri}, \citenamefont {Weihs},\ and\ \citenamefont
  {Zeilinger}}]{Mair2001}%
  \BibitemOpen
  \bibfield  {author} {\bibinfo {author} {\bibfnamefont {A.}~\bibnamefont
  {Mair}}, \bibinfo {author} {\bibfnamefont {A.}~\bibnamefont {Vaziri}},
  \bibinfo {author} {\bibfnamefont {G.}~\bibnamefont {Weihs}},\ and\ \bibinfo
  {author} {\bibfnamefont {A.}~\bibnamefont {Zeilinger}},\ }\href
  {https://doi.org/10.1038/35085529} {\bibfield  {journal} {\bibinfo  {journal}
  {Nature}\ }\textbf {\bibinfo {volume} {412}},\ \bibinfo {pages} {313}
  (\bibinfo {year} {2001})}\BibitemShut {NoStop}%
\bibitem [{\citenamefont {Franke-Arnold}\ \emph {et~al.}(2002)\citenamefont
  {Franke-Arnold}, \citenamefont {Barnett}, \citenamefont {Padgett},\ and\
  \citenamefont {Allen}}]{Franke-arnold2002b}%
  \BibitemOpen
  \bibfield  {author} {\bibinfo {author} {\bibfnamefont {S.}~\bibnamefont
  {Franke-Arnold}}, \bibinfo {author} {\bibfnamefont {S.~M.}\ \bibnamefont
  {Barnett}}, \bibinfo {author} {\bibfnamefont {M.~J.}\ \bibnamefont
  {Padgett}},\ and\ \bibinfo {author} {\bibfnamefont {L.}~\bibnamefont
  {Allen}},\ }\href {https://doi.org/10.1103/PhysRevA.65.033823} {\bibfield
  {journal} {\bibinfo  {journal} {Phys. Rev. A}\ }\textbf {\bibinfo {volume}
  {65}},\ \bibinfo {pages} {033823} (\bibinfo {year} {2002})}\BibitemShut
  {NoStop}%
\bibitem [{\citenamefont {Walker}\ \emph {et~al.}(2012)\citenamefont {Walker},
  \citenamefont {Arnold},\ and\ \citenamefont {Franke-Arnold}}]{Walker2012}%
  \BibitemOpen
  \bibfield  {author} {\bibinfo {author} {\bibfnamefont {G.}~\bibnamefont
  {Walker}}, \bibinfo {author} {\bibfnamefont {A.~S.}\ \bibnamefont {Arnold}},\
  and\ \bibinfo {author} {\bibfnamefont {S.}~\bibnamefont {Franke-Arnold}},\
  }\href {https://doi.org/10.1103/PhysRevLett.108.243601} {\bibfield  {journal}
  {\bibinfo  {journal} {Phys. Rev. Lett.}\ }\textbf {\bibinfo {volume} {108}},\
  \bibinfo {pages} {243601} (\bibinfo {year} {2012})}\BibitemShut {NoStop}%
\bibitem [{\citenamefont {Akulshin}\ \emph {et~al.}(2016)\citenamefont
  {Akulshin}, \citenamefont {Novikova}, \citenamefont {Mikhailov},
  \citenamefont {Suslov},\ and\ \citenamefont {McLean}}]{Akulshin2016}%
  \BibitemOpen
  \bibfield  {author} {\bibinfo {author} {\bibfnamefont {A.~M.}\ \bibnamefont
  {Akulshin}}, \bibinfo {author} {\bibfnamefont {I.}~\bibnamefont {Novikova}},
  \bibinfo {author} {\bibfnamefont {E.~E.}\ \bibnamefont {Mikhailov}}, \bibinfo
  {author} {\bibfnamefont {S.~A.}\ \bibnamefont {Suslov}},\ and\ \bibinfo
  {author} {\bibfnamefont {R.~J.}\ \bibnamefont {McLean}},\ }\href
  {https://doi.org/10.1364/ol.41.001146} {\bibfield  {journal} {\bibinfo
  {journal} {Opt.\ Lett.}\ }\textbf {\bibinfo {volume} {41}},\ \bibinfo {pages}
  {1146} (\bibinfo {year} {2016})}\BibitemShut {NoStop}%
\bibitem [{\citenamefont {Chopinaud}\ \emph {et~al.}(2018)\citenamefont
  {Chopinaud}, \citenamefont {Jacquey}, \citenamefont {{Viaris de Lesegno}},\
  and\ \citenamefont {Pruvost}}]{Chopinaud2018}%
  \BibitemOpen
  \bibfield  {author} {\bibinfo {author} {\bibfnamefont {A.}~\bibnamefont
  {Chopinaud}}, \bibinfo {author} {\bibfnamefont {M.}~\bibnamefont {Jacquey}},
  \bibinfo {author} {\bibfnamefont {B.}~\bibnamefont {{Viaris de Lesegno}}},\
  and\ \bibinfo {author} {\bibfnamefont {L.}~\bibnamefont {Pruvost}},\ }\href
  {https://doi.org/10.1103/PhysRevA.97.063806} {\bibfield  {journal} {\bibinfo
  {journal} {Phys. Rev. A}\ }\textbf {\bibinfo {volume} {97}},\ \bibinfo
  {pages} {063806} (\bibinfo {year} {2018})}\BibitemShut {NoStop}%
\bibitem [{\citenamefont {Offer}\ \emph {et~al.}(2018)\citenamefont {Offer},
  \citenamefont {Stulga}, \citenamefont {Riis}, \citenamefont {Franke-Arnold},\
  and\ \citenamefont {Arnold}}]{Offer2018}%
  \BibitemOpen
  \bibfield  {author} {\bibinfo {author} {\bibfnamefont {R.~F.}\ \bibnamefont
  {Offer}}, \bibinfo {author} {\bibfnamefont {D.}~\bibnamefont {Stulga}},
  \bibinfo {author} {\bibfnamefont {E.}~\bibnamefont {Riis}}, \bibinfo {author}
  {\bibfnamefont {S.}~\bibnamefont {Franke-Arnold}},\ and\ \bibinfo {author}
  {\bibfnamefont {A.~S.}\ \bibnamefont {Arnold}},\ }\href
  {https://doi.org/10.1038/s42005-018-0077-5} {\bibfield  {journal} {\bibinfo
  {journal} {Commun.\ Phys.}\ }\textbf {\bibinfo {volume} {1}},\ \bibinfo
  {pages} {84} (\bibinfo {year} {2018})}\BibitemShut {NoStop}%
\bibitem [{\citenamefont {Pereira}\ \emph {et~al.}(2017)\citenamefont
  {Pereira}, \citenamefont {Buono}, \citenamefont {Tasca}, \citenamefont
  {Dechoum},\ and\ \citenamefont {Khoury}}]{Khoury2017}%
  \BibitemOpen
  \bibfield  {author} {\bibinfo {author} {\bibfnamefont {L.~J.}\ \bibnamefont
  {Pereira}}, \bibinfo {author} {\bibfnamefont {W.~T.}\ \bibnamefont {Buono}},
  \bibinfo {author} {\bibfnamefont {D.~S.}\ \bibnamefont {Tasca}}, \bibinfo
  {author} {\bibfnamefont {K.}~\bibnamefont {Dechoum}},\ and\ \bibinfo {author}
  {\bibfnamefont {A.~Z.}\ \bibnamefont {Khoury}},\ }\href
  {https://doi.org/10.1103/PhysRevA.96.053856} {\bibfield  {journal} {\bibinfo
  {journal} {Phys.\ Rev.\ A}\ }\textbf {\bibinfo {volume} {96}},\ \bibinfo
  {pages} {053856} (\bibinfo {year} {2017})}\BibitemShut {NoStop}%
\bibitem [{\citenamefont {Alves}\ \emph {et~al.}(2018)\citenamefont {Alves},
  \citenamefont {Barros}, \citenamefont {Tasca}, \citenamefont {Souza},\ and\
  \citenamefont {Khoury}}]{Khoury2018}%
  \BibitemOpen
  \bibfield  {author} {\bibinfo {author} {\bibfnamefont {G.~B.}\ \bibnamefont
  {Alves}}, \bibinfo {author} {\bibfnamefont {R.~F.}\ \bibnamefont {Barros}},
  \bibinfo {author} {\bibfnamefont {D.~S.}\ \bibnamefont {Tasca}}, \bibinfo
  {author} {\bibfnamefont {C.~E.~R.}\ \bibnamefont {Souza}},\ and\ \bibinfo
  {author} {\bibfnamefont {A.~Z.}\ \bibnamefont {Khoury}},\ }\href
  {https://doi.org/10.1103/physreva.98.063825} {\bibfield  {journal} {\bibinfo
  {journal} {Phys.\ Rev.\ A}\ }\textbf {\bibinfo {volume} {98}},\ \bibinfo
  {pages} {063825} (\bibinfo {year} {2018})}\BibitemShut {NoStop}%
\bibitem [{\citenamefont {da~Silva}\ \emph {et~al.}(2019)\citenamefont
  {da~Silva}, \citenamefont {Tasca}, \citenamefont {Galv{\~{a}}o},\ and\
  \citenamefont {Khoury}}]{Khoury2019}%
  \BibitemOpen
  \bibfield  {author} {\bibinfo {author} {\bibfnamefont {B.~P.}\ \bibnamefont
  {da~Silva}}, \bibinfo {author} {\bibfnamefont {D.~S.}\ \bibnamefont {Tasca}},
  \bibinfo {author} {\bibfnamefont {E.~F.}\ \bibnamefont {Galv{\~{a}}o}},\ and\
  \bibinfo {author} {\bibfnamefont {A.~Z.}\ \bibnamefont {Khoury}},\ }\href
  {https://doi.org/10.1103/physreva.99.043820} {\bibfield  {journal} {\bibinfo
  {journal} {Phys.\ Rev.\ A}\ }\textbf {\bibinfo {volume} {99}},\ \bibinfo
  {pages} {043820} (\bibinfo {year} {2019})}\BibitemShut {NoStop}%
\bibitem [{\citenamefont {Wu}\ \emph {et~al.}(2020)\citenamefont {Wu},
  \citenamefont {Mao}, \citenamefont {Yang}, \citenamefont
  {Rosales-Guzm{\'{a}}n}, \citenamefont {Gao}, \citenamefont {Shi},\ and\
  \citenamefont {Zhu}}]{Zhu2020}%
  \BibitemOpen
  \bibfield  {author} {\bibinfo {author} {\bibfnamefont {H.-J.}\ \bibnamefont
  {Wu}}, \bibinfo {author} {\bibfnamefont {L.-W.}\ \bibnamefont {Mao}},
  \bibinfo {author} {\bibfnamefont {Y.-J.}\ \bibnamefont {Yang}}, \bibinfo
  {author} {\bibfnamefont {C.}~\bibnamefont {Rosales-Guzm{\'{a}}n}}, \bibinfo
  {author} {\bibfnamefont {W.}~\bibnamefont {Gao}}, \bibinfo {author}
  {\bibfnamefont {B.-S.}\ \bibnamefont {Shi}},\ and\ \bibinfo {author}
  {\bibfnamefont {Z.-H.}\ \bibnamefont {Zhu}},\ }\href
  {https://doi.org/10.1103/physreva.101.063805} {\bibfield  {journal} {\bibinfo
   {journal} {Phys.\ Rev.\ A}\ }\textbf {\bibinfo {volume} {101}},\ \bibinfo
  {pages} {063805} (\bibinfo {year} {2020})}\BibitemShut {NoStop}%
\bibitem [{\citenamefont {Buono}\ \emph {et~al.}(2020)\citenamefont {Buono},
  \citenamefont {Santos}, \citenamefont {Maia}, \citenamefont {Pereira},
  \citenamefont {Tasca}, \citenamefont {Dechoum}, \citenamefont {Ruchon},\ and\
  \citenamefont {Khoury}}]{Khoury2020}%
  \BibitemOpen
  \bibfield  {author} {\bibinfo {author} {\bibfnamefont {W.~T.}\ \bibnamefont
  {Buono}}, \bibinfo {author} {\bibfnamefont {A.}~\bibnamefont {Santos}},
  \bibinfo {author} {\bibfnamefont {M.~R.}\ \bibnamefont {Maia}}, \bibinfo
  {author} {\bibfnamefont {L.~J.}\ \bibnamefont {Pereira}}, \bibinfo {author}
  {\bibfnamefont {D.~S.}\ \bibnamefont {Tasca}}, \bibinfo {author}
  {\bibfnamefont {K.}~\bibnamefont {Dechoum}}, \bibinfo {author} {\bibfnamefont
  {T.}~\bibnamefont {Ruchon}},\ and\ \bibinfo {author} {\bibfnamefont {A.~Z.}\
  \bibnamefont {Khoury}},\ }\href {https://doi.org/10.1103/physreva.101.043821}
  {\bibfield  {journal} {\bibinfo  {journal} {Phys.\ Rev.\ A}\ }\textbf
  {\bibinfo {volume} {101}},\ \bibinfo {pages} {043821} (\bibinfo {year}
  {2020})}\BibitemShut {NoStop}%
\bibitem [{\citenamefont {Zhao}\ \emph {et~al.}(2020)\citenamefont {Zhao},
  \citenamefont {Lu}, \citenamefont {Hong}, \citenamefont {Feng}, \citenamefont
  {Xu}, \citenamefont {Yuan}, \citenamefont {Zhang}, \citenamefont {Qin},\ and\
  \citenamefont {Zhu}}]{Zhao2020}%
  \BibitemOpen
  \bibfield  {author} {\bibinfo {author} {\bibfnamefont {R.-Z.}\ \bibnamefont
  {Zhao}}, \bibinfo {author} {\bibfnamefont {R.-E.}\ \bibnamefont {Lu}},
  \bibinfo {author} {\bibfnamefont {X.-H.}\ \bibnamefont {Hong}}, \bibinfo
  {author} {\bibfnamefont {X.}~\bibnamefont {Feng}}, \bibinfo {author}
  {\bibfnamefont {Y.-G.}\ \bibnamefont {Xu}}, \bibinfo {author} {\bibfnamefont
  {X.-D.}\ \bibnamefont {Yuan}}, \bibinfo {author} {\bibfnamefont
  {C.}~\bibnamefont {Zhang}}, \bibinfo {author} {\bibfnamefont {Y.-Q.}\
  \bibnamefont {Qin}},\ and\ \bibinfo {author} {\bibfnamefont {Y.-Y.}\
  \bibnamefont {Zhu}},\ }\href {https://doi.org/10.1103/physreva.101.023834}
  {\bibfield  {journal} {\bibinfo  {journal} {Phys.\ Rev.\ A}\ }\textbf
  {\bibinfo {volume} {101}},\ \bibinfo {pages} {023834} (\bibinfo {year}
  {2020})}\BibitemShut {NoStop}%
\bibitem [{\citenamefont {Lanning}\ \emph {et~al.}(2017)\citenamefont
  {Lanning}, \citenamefont {Xiao}, \citenamefont {Zhang}, \citenamefont
  {Novikova}, \citenamefont {Mikhailov},\ and\ \citenamefont
  {Dowling}}]{Lanning2017}%
  \BibitemOpen
  \bibfield  {author} {\bibinfo {author} {\bibfnamefont {R.~N.}\ \bibnamefont
  {Lanning}}, \bibinfo {author} {\bibfnamefont {Z.}~\bibnamefont {Xiao}},
  \bibinfo {author} {\bibfnamefont {M.}~\bibnamefont {Zhang}}, \bibinfo
  {author} {\bibfnamefont {I.}~\bibnamefont {Novikova}}, \bibinfo {author}
  {\bibfnamefont {E.~E.}\ \bibnamefont {Mikhailov}},\ and\ \bibinfo {author}
  {\bibfnamefont {J.~P.}\ \bibnamefont {Dowling}},\ }\href
  {https://doi.org/10.1103/PhysRevA.96.013830} {\bibfield  {journal} {\bibinfo
  {journal} {Phys. Rev. A}\ }\textbf {\bibinfo {volume} {96}},\ \bibinfo
  {pages} {013830} (\bibinfo {year} {2017})}\BibitemShut {NoStop}%
\bibitem [{\citenamefont {Boyd}\ and\ \citenamefont
  {Kleinman}(1968)}]{Boyd1968}%
  \BibitemOpen
  \bibfield  {author} {\bibinfo {author} {\bibfnamefont {G.~D.}\ \bibnamefont
  {Boyd}}\ and\ \bibinfo {author} {\bibfnamefont {D.~A.}\ \bibnamefont
  {Kleinman}},\ }\href {https://doi.org/10.1063/1.1656831} {\bibfield
  {journal} {\bibinfo  {journal} {J. Appl. Phys.}\ }\textbf {\bibinfo {volume}
  {39}},\ \bibinfo {pages} {3597} (\bibinfo {year} {1968})}\BibitemShut
  {NoStop}%
\bibitem [{\citenamefont {Paterson}\ \emph {et~al.}(2001)\citenamefont
  {Paterson}, \citenamefont {MacDonald}, \citenamefont {Arlt}, \citenamefont
  {Sibbett}, \citenamefont {Bryant},\ and\ \citenamefont
  {Dholakia}}]{Paterson2001}%
  \BibitemOpen
  \bibfield  {author} {\bibinfo {author} {\bibfnamefont {L.}~\bibnamefont
  {Paterson}}, \bibinfo {author} {\bibfnamefont {M.~P.}\ \bibnamefont
  {MacDonald}}, \bibinfo {author} {\bibfnamefont {J.}~\bibnamefont {Arlt}},
  \bibinfo {author} {\bibfnamefont {W.}~\bibnamefont {Sibbett}}, \bibinfo
  {author} {\bibfnamefont {P.~E.}\ \bibnamefont {Bryant}},\ and\ \bibinfo
  {author} {\bibfnamefont {K.}~\bibnamefont {Dholakia}},\ }\href
  {https://doi.org/10.1126/science.1058591} {\bibfield  {journal} {\bibinfo
  {journal} {Science}\ }\textbf {\bibinfo {volume} {292}},\ \bibinfo {pages}
  {912} (\bibinfo {year} {2001})}\BibitemShut {NoStop}%
\bibitem [{\citenamefont {Amico}\ \emph {et~al.}(2005)\citenamefont {Amico},
  \citenamefont {Osterloh},\ and\ \citenamefont {Cataliotti}}]{Amico2005}%
  \BibitemOpen
  \bibfield  {author} {\bibinfo {author} {\bibfnamefont {L.}~\bibnamefont
  {Amico}}, \bibinfo {author} {\bibfnamefont {A.}~\bibnamefont {Osterloh}},\
  and\ \bibinfo {author} {\bibfnamefont {F.}~\bibnamefont {Cataliotti}},\
  }\href {https://doi.org/10.1103/physrevlett.95.063201} {\bibfield  {journal}
  {\bibinfo  {journal} {Phys.\ Rev.\ Lett.}\ }\textbf {\bibinfo {volume}
  {95}},\ \bibinfo {pages} {063201} (\bibinfo {year} {2005})}\BibitemShut
  {NoStop}%
\bibitem [{\citenamefont {Franke-Arnold}\ \emph {et~al.}(2007)\citenamefont
  {Franke-Arnold}, \citenamefont {Leach}, \citenamefont {Padgett},
  \citenamefont {Lembessis}, \citenamefont {Ellinas}, \citenamefont {Wright},
  \citenamefont {Girkin}, \citenamefont {\"{O}hberg},\ and\ \citenamefont
  {Arnold}}]{Franke-Arnold2007a}%
  \BibitemOpen
  \bibfield  {author} {\bibinfo {author} {\bibfnamefont {S.}~\bibnamefont
  {Franke-Arnold}}, \bibinfo {author} {\bibfnamefont {J.}~\bibnamefont
  {Leach}}, \bibinfo {author} {\bibfnamefont {M.~J.}\ \bibnamefont {Padgett}},
  \bibinfo {author} {\bibfnamefont {V.~E.}\ \bibnamefont {Lembessis}}, \bibinfo
  {author} {\bibfnamefont {D.}~\bibnamefont {Ellinas}}, \bibinfo {author}
  {\bibfnamefont {A.~J.}\ \bibnamefont {Wright}}, \bibinfo {author}
  {\bibfnamefont {J.~M.}\ \bibnamefont {Girkin}}, \bibinfo {author}
  {\bibfnamefont {P.}~\bibnamefont {\"{O}hberg}},\ and\ \bibinfo {author}
  {\bibfnamefont {A.~S.}\ \bibnamefont {Arnold}},\ }\href
  {https://doi.org/10.1364/OE.15.008619} {\bibfield  {journal} {\bibinfo
  {journal} {Opt.\ Express}\ }\textbf {\bibinfo {volume} {15}},\ \bibinfo
  {pages} {8619} (\bibinfo {year} {2007})}\BibitemShut {NoStop}%
\bibitem [{\citenamefont {Bhattacharya}(2007)}]{Bhattacharya2007}%
  \BibitemOpen
  \bibfield  {author} {\bibinfo {author} {\bibfnamefont {M.}~\bibnamefont
  {Bhattacharya}},\ }\href {https://doi.org/10.1016/j.optcom.2007.07.008}
  {\bibfield  {journal} {\bibinfo  {journal} {Opt.\ Commun.}\ }\textbf
  {\bibinfo {volume} {279}},\ \bibinfo {pages} {219} (\bibinfo {year}
  {2007})}\BibitemShut {NoStop}%
\bibitem [{\citenamefont {Arnold}(2012)}]{Arnold2012}%
  \BibitemOpen
  \bibfield  {author} {\bibinfo {author} {\bibfnamefont {A.~S.}\ \bibnamefont
  {Arnold}},\ }\href {https://doi.org/10.1364/ol.37.002505} {\bibfield
  {journal} {\bibinfo  {journal} {Opt.\ Lett.}\ }\textbf {\bibinfo {volume}
  {37}},\ \bibinfo {pages} {2505} (\bibinfo {year} {2012})}\BibitemShut
  {NoStop}%
\bibitem [{\citenamefont {Alperin}\ \emph {et~al.}(2016)\citenamefont
  {Alperin}, \citenamefont {Niederriter}, \citenamefont {Gopinath},\ and\
  \citenamefont {Siemens}}]{Siemens2016}%
  \BibitemOpen
  \bibfield  {author} {\bibinfo {author} {\bibfnamefont {S.~N.}\ \bibnamefont
  {Alperin}}, \bibinfo {author} {\bibfnamefont {R.~D.}\ \bibnamefont
  {Niederriter}}, \bibinfo {author} {\bibfnamefont {J.~T.}\ \bibnamefont
  {Gopinath}},\ and\ \bibinfo {author} {\bibfnamefont {M.~E.}\ \bibnamefont
  {Siemens}},\ }\href {https://doi.org/10.1364/ol.41.005019} {\bibfield
  {journal} {\bibinfo  {journal} {Opt.\ Lett.}\ }\textbf {\bibinfo {volume}
  {41}},\ \bibinfo {pages} {5019} (\bibinfo {year} {2016})}\BibitemShut
  {NoStop}%
\bibitem [{\citenamefont {Zibrov}\ \emph {et~al.}(2002)\citenamefont {Zibrov},
  \citenamefont {Lukin}, \citenamefont {Hollberg},\ and\ \citenamefont
  {Scully}}]{Zibrov2002}%
  \BibitemOpen
  \bibfield  {author} {\bibinfo {author} {\bibfnamefont {A.~S.}\ \bibnamefont
  {Zibrov}}, \bibinfo {author} {\bibfnamefont {M.~D.}\ \bibnamefont {Lukin}},
  \bibinfo {author} {\bibfnamefont {L.}~\bibnamefont {Hollberg}},\ and\
  \bibinfo {author} {\bibfnamefont {M.~O.}\ \bibnamefont {Scully}},\ }\href
  {https://doi.org/10.1103/PhysRevA.65.051801} {\bibfield  {journal} {\bibinfo
  {journal} {Phys. Rev. A}\ }\textbf {\bibinfo {volume} {65}},\ \bibinfo
  {pages} {051801(R)} (\bibinfo {year} {2002})}\BibitemShut {NoStop}%
\bibitem [{\citenamefont {Meijer}\ \emph {et~al.}(2006)\citenamefont {Meijer},
  \citenamefont {White}, \citenamefont {Smeets}, \citenamefont {Jeppesen},\
  and\ \citenamefont {Scholten}}]{Meijer2006}%
  \BibitemOpen
  \bibfield  {author} {\bibinfo {author} {\bibfnamefont {T.}~\bibnamefont
  {Meijer}}, \bibinfo {author} {\bibfnamefont {J.~D.}\ \bibnamefont {White}},
  \bibinfo {author} {\bibfnamefont {B.}~\bibnamefont {Smeets}}, \bibinfo
  {author} {\bibfnamefont {M.}~\bibnamefont {Jeppesen}},\ and\ \bibinfo
  {author} {\bibfnamefont {R.~E.}\ \bibnamefont {Scholten}},\ }\href
  {https://doi.org/10.1364/OL.31.001002} {\bibfield  {journal} {\bibinfo
  {journal} {Opt. Lett.}\ }\textbf {\bibinfo {volume} {31}},\ \bibinfo {pages}
  {1002} (\bibinfo {year} {2006})}\BibitemShut {NoStop}%
\bibitem [{\citenamefont {Schultz}\ \emph {et~al.}(2009)\citenamefont
  {Schultz}, \citenamefont {Abend}, \citenamefont {D{\"{o}}ring}, \citenamefont
  {Debs}, \citenamefont {Altin}, \citenamefont {White}, \citenamefont
  {Robins},\ and\ \citenamefont {Close}}]{Schultz2009}%
  \BibitemOpen
  \bibfield  {author} {\bibinfo {author} {\bibfnamefont {J.~T.}\ \bibnamefont
  {Schultz}}, \bibinfo {author} {\bibfnamefont {S.}~\bibnamefont {Abend}},
  \bibinfo {author} {\bibfnamefont {D.}~\bibnamefont {D{\"{o}}ring}}, \bibinfo
  {author} {\bibfnamefont {J.~E.}\ \bibnamefont {Debs}}, \bibinfo {author}
  {\bibfnamefont {P.~A.}\ \bibnamefont {Altin}}, \bibinfo {author}
  {\bibfnamefont {J.~D.}\ \bibnamefont {White}}, \bibinfo {author}
  {\bibfnamefont {N.~P.}\ \bibnamefont {Robins}},\ and\ \bibinfo {author}
  {\bibfnamefont {J.~D.}\ \bibnamefont {Close}},\ }\href
  {https://doi.org/10.1364/OL.34.002321} {\bibfield  {journal} {\bibinfo
  {journal} {Opt. Lett.}\ }\textbf {\bibinfo {volume} {34}},\ \bibinfo {pages}
  {2321} (\bibinfo {year} {2009})}\BibitemShut {NoStop}%
\bibitem [{\citenamefont {Akulshin}\ \emph {et~al.}(2009)\citenamefont
  {Akulshin}, \citenamefont {McLean}, \citenamefont {Sidorov},\ and\
  \citenamefont {Hannaford}}]{Akulshin2009}%
  \BibitemOpen
  \bibfield  {author} {\bibinfo {author} {\bibfnamefont {A.~M.}\ \bibnamefont
  {Akulshin}}, \bibinfo {author} {\bibfnamefont {R.~J.}\ \bibnamefont
  {McLean}}, \bibinfo {author} {\bibfnamefont {A.~I.}\ \bibnamefont
  {Sidorov}},\ and\ \bibinfo {author} {\bibfnamefont {P.}~\bibnamefont
  {Hannaford}},\ }\href {https://doi.org/10.1364/OE.17.022861} {\bibfield
  {journal} {\bibinfo  {journal} {Opt. Express}\ }\textbf {\bibinfo {volume}
  {17}},\ \bibinfo {pages} {22861} (\bibinfo {year} {2009})}\BibitemShut
  {NoStop}%
\bibitem [{\citenamefont {Vernier}\ \emph {et~al.}(2010)\citenamefont
  {Vernier}, \citenamefont {Franke-Arnold}, \citenamefont {Riis},\ and\
  \citenamefont {Arnold}}]{Vernier2010}%
  \BibitemOpen
  \bibfield  {author} {\bibinfo {author} {\bibfnamefont {A.}~\bibnamefont
  {Vernier}}, \bibinfo {author} {\bibfnamefont {S.}~\bibnamefont
  {Franke-Arnold}}, \bibinfo {author} {\bibfnamefont {E.}~\bibnamefont
  {Riis}},\ and\ \bibinfo {author} {\bibfnamefont {A.~S.}\ \bibnamefont
  {Arnold}},\ }\href {https://doi.org/10.1364/OE.18.017020} {\bibfield
  {journal} {\bibinfo  {journal} {Opt. Express}\ }\textbf {\bibinfo {volume}
  {18}},\ \bibinfo {pages} {17020} (\bibinfo {year} {2010})}\BibitemShut
  {NoStop}%
\bibitem [{\citenamefont {Offer}\ \emph {et~al.}(2016)\citenamefont {Offer},
  \citenamefont {Conway}, \citenamefont {Riis}, \citenamefont {Franke-Arnold},\
  and\ \citenamefont {Arnold}}]{Offer2016}%
  \BibitemOpen
  \bibfield  {author} {\bibinfo {author} {\bibfnamefont {R.~F.}\ \bibnamefont
  {Offer}}, \bibinfo {author} {\bibfnamefont {J.~W.~C.}\ \bibnamefont
  {Conway}}, \bibinfo {author} {\bibfnamefont {E.}~\bibnamefont {Riis}},
  \bibinfo {author} {\bibfnamefont {S.}~\bibnamefont {Franke-Arnold}},\ and\
  \bibinfo {author} {\bibfnamefont {A.~S.}\ \bibnamefont {Arnold}},\ }\href
  {https://doi.org/10.1364/ol.41.002177} {\bibfield  {journal} {\bibinfo
  {journal} {Opt.\ Lett.}\ }\textbf {\bibinfo {volume} {41}},\ \bibinfo {pages}
  {2177} (\bibinfo {year} {2016})}\BibitemShut {NoStop}%
\bibitem [{\citenamefont {Brekke}\ and\ \citenamefont
  {Potier}(2017)}]{Brekke2016}%
  \BibitemOpen
  \bibfield  {author} {\bibinfo {author} {\bibfnamefont {E.}~\bibnamefont
  {Brekke}}\ and\ \bibinfo {author} {\bibfnamefont {S.}~\bibnamefont
  {Potier}},\ }\href {https://doi.org/10.1364/AO.56.000046} {\bibfield
  {journal} {\bibinfo  {journal} {Appl. Opt.}\ }\textbf {\bibinfo {volume}
  {56}},\ \bibinfo {pages} {46} (\bibinfo {year} {2017})}\BibitemShut {NoStop}%
\bibitem [{\citenamefont {Arnold}\ \emph {et~al.}(1998)\citenamefont {Arnold},
  \citenamefont {Wilson},\ and\ \citenamefont {Boshier}}]{Arnold1998}%
  \BibitemOpen
  \bibfield  {author} {\bibinfo {author} {\bibfnamefont {A.~S.}\ \bibnamefont
  {Arnold}}, \bibinfo {author} {\bibfnamefont {J.~S.}\ \bibnamefont {Wilson}},\
  and\ \bibinfo {author} {\bibfnamefont {M.~G.}\ \bibnamefont {Boshier}},\
  }\href {https://doi.org/10.1063/1.1148756} {\bibfield  {journal} {\bibinfo
  {journal} {Rev. Sci. Instrum.}\ }\textbf {\bibinfo {volume} {69}},\ \bibinfo
  {pages} {1236} (\bibinfo {year} {1998})}\BibitemShut {NoStop}%
\bibitem [{\citenamefont {Clark}\ \emph {et~al.}(2016)\citenamefont {Clark},
  \citenamefont {Offer}, \citenamefont {Franke-Arnold}, \citenamefont
  {Arnold},\ and\ \citenamefont {Radwell}}]{Clark2016}%
  \BibitemOpen
  \bibfield  {author} {\bibinfo {author} {\bibfnamefont {T.~W.}\ \bibnamefont
  {Clark}}, \bibinfo {author} {\bibfnamefont {R.~F.}\ \bibnamefont {Offer}},
  \bibinfo {author} {\bibfnamefont {S.}~\bibnamefont {Franke-Arnold}}, \bibinfo
  {author} {\bibfnamefont {A.~S.}\ \bibnamefont {Arnold}},\ and\ \bibinfo
  {author} {\bibfnamefont {N.}~\bibnamefont {Radwell}},\ }\href
  {https://doi.org/10.1364/OE.24.006249} {\bibfield  {journal} {\bibinfo
  {journal} {Opt. Express}\ }\textbf {\bibinfo {volume} {24}},\ \bibinfo
  {pages} {6249} (\bibinfo {year} {2016})}\BibitemShut {NoStop}%
\bibitem [{\citenamefont {Wang}\ \emph {et~al.}(2019)\citenamefont {Wang},
  \citenamefont {Sciarrino}, \citenamefont {Laing},\ and\ \citenamefont
  {Thompson}}]{Wang2019}%
  \BibitemOpen
  \bibfield  {author} {\bibinfo {author} {\bibfnamefont {J.}~\bibnamefont
  {Wang}}, \bibinfo {author} {\bibfnamefont {F.}~\bibnamefont {Sciarrino}},
  \bibinfo {author} {\bibfnamefont {A.}~\bibnamefont {Laing}},\ and\ \bibinfo
  {author} {\bibfnamefont {M.~G.}\ \bibnamefont {Thompson}},\ }\href
  {https://doi.org/10.1038/s41566-019-0532-1} {\bibfield  {journal} {\bibinfo
  {journal} {Nat.\ Photonics}\ }\textbf {\bibinfo {volume} {14}},\ \bibinfo
  {pages} {273} (\bibinfo {year} {2019})}\BibitemShut {NoStop}%
\bibitem [{\citenamefont {Kim}\ \emph {et~al.}(2020)\citenamefont {Kim},
  \citenamefont {Aghaeimeibodi}, \citenamefont {Carolan}, \citenamefont
  {Englund},\ and\ \citenamefont {Waks}}]{Kim2020}%
  \BibitemOpen
  \bibfield  {author} {\bibinfo {author} {\bibfnamefont {J.-H.}\ \bibnamefont
  {Kim}}, \bibinfo {author} {\bibfnamefont {S.}~\bibnamefont {Aghaeimeibodi}},
  \bibinfo {author} {\bibfnamefont {J.}~\bibnamefont {Carolan}}, \bibinfo
  {author} {\bibfnamefont {D.}~\bibnamefont {Englund}},\ and\ \bibinfo {author}
  {\bibfnamefont {E.}~\bibnamefont {Waks}},\ }\href
  {https://doi.org/10.1364/optica.384118} {\bibfield  {journal} {\bibinfo
  {journal} {Optica}\ }\textbf {\bibinfo {volume} {7}},\ \bibinfo {pages} {291}
  (\bibinfo {year} {2020})}\BibitemShut {NoStop}%
\bibitem [{\citenamefont {Chen}\ \emph {et~al.}(2018)\citenamefont {Chen},
  \citenamefont {Gao}, \citenamefont {Jiao}, \citenamefont {Sun}, \citenamefont
  {Shen}, \citenamefont {Qiao}, \citenamefont {Tang}, \citenamefont {Lin},\
  and\ \citenamefont {Jin}}]{Chen20188}%
  \BibitemOpen
  \bibfield  {author} {\bibinfo {author} {\bibfnamefont {Y.}~\bibnamefont
  {Chen}}, \bibinfo {author} {\bibfnamefont {J.}~\bibnamefont {Gao}}, \bibinfo
  {author} {\bibfnamefont {Z.-Q.}\ \bibnamefont {Jiao}}, \bibinfo {author}
  {\bibfnamefont {K.}~\bibnamefont {Sun}}, \bibinfo {author} {\bibfnamefont
  {W.-G.}\ \bibnamefont {Shen}}, \bibinfo {author} {\bibfnamefont {L.-F.}\
  \bibnamefont {Qiao}}, \bibinfo {author} {\bibfnamefont {H.}~\bibnamefont
  {Tang}}, \bibinfo {author} {\bibfnamefont {X.-F.}\ \bibnamefont {Lin}},\ and\
  \bibinfo {author} {\bibfnamefont {X.-M.}\ \bibnamefont {Jin}},\ }\href
  {https://doi.org/10.1103/physrevlett.121.233602} {\bibfield  {journal}
  {\bibinfo  {journal} {Phys.\ Rev.\ Lett.}\ }\textbf {\bibinfo {volume}
  {121}},\ \bibinfo {pages} {233602} (\bibinfo {year} {2018})}\BibitemShut
  {NoStop}%
\bibitem [{\citenamefont {Chen}\ \emph {et~al.}(2020)\citenamefont {Chen},
  \citenamefont {Xia}, \citenamefont {Shen}, \citenamefont {Gao}, \citenamefont
  {Yan}, \citenamefont {Jiao}, \citenamefont {Dou}, \citenamefont {Tang},
  \citenamefont {Lu},\ and\ \citenamefont {Jin}}]{Chen2020}%
  \BibitemOpen
  \bibfield  {author} {\bibinfo {author} {\bibfnamefont {Y.}~\bibnamefont
  {Chen}}, \bibinfo {author} {\bibfnamefont {K.-Y.}\ \bibnamefont {Xia}},
  \bibinfo {author} {\bibfnamefont {W.-G.}\ \bibnamefont {Shen}}, \bibinfo
  {author} {\bibfnamefont {J.}~\bibnamefont {Gao}}, \bibinfo {author}
  {\bibfnamefont {Z.-Q.}\ \bibnamefont {Yan}}, \bibinfo {author} {\bibfnamefont
  {Z.-Q.}\ \bibnamefont {Jiao}}, \bibinfo {author} {\bibfnamefont {J.-P.}\
  \bibnamefont {Dou}}, \bibinfo {author} {\bibfnamefont {H.}~\bibnamefont
  {Tang}}, \bibinfo {author} {\bibfnamefont {Y.-Q.}\ \bibnamefont {Lu}},\ and\
  \bibinfo {author} {\bibfnamefont {X.-M.}\ \bibnamefont {Jin}},\ }\href
  {https://doi.org/10.1103/physrevlett.124.153601} {\bibfield  {journal}
  {\bibinfo  {journal} {Phys.\ Rev.\ Lett.}\ }\textbf {\bibinfo {volume}
  {124}},\ \bibinfo {pages} {153601} (\bibinfo {year} {2020})}\BibitemShut
  {NoStop}%
\bibitem [{\citenamefont {Cutler}\ \emph {et~al.}(2020)\citenamefont {Cutler},
  \citenamefont {Hamlyn}, \citenamefont {Renger}, \citenamefont {Whittaker},
  \citenamefont {Pizzey}, \citenamefont {Hughes}, \citenamefont {Sandoghdar},\
  and\ \citenamefont {Adams}}]{Adams2020}%
  \BibitemOpen
  \bibfield  {author} {\bibinfo {author} {\bibfnamefont {T.~F.}\ \bibnamefont
  {Cutler}}, \bibinfo {author} {\bibfnamefont {W.~J.}\ \bibnamefont {Hamlyn}},
  \bibinfo {author} {\bibfnamefont {J.}~\bibnamefont {Renger}}, \bibinfo
  {author} {\bibfnamefont {K.~A.}\ \bibnamefont {Whittaker}}, \bibinfo {author}
  {\bibfnamefont {D.}~\bibnamefont {Pizzey}}, \bibinfo {author} {\bibfnamefont
  {I.~G.}\ \bibnamefont {Hughes}}, \bibinfo {author} {\bibfnamefont
  {V.}~\bibnamefont {Sandoghdar}},\ and\ \bibinfo {author} {\bibfnamefont
  {C.~S.}\ \bibnamefont {Adams}},\ }\href
  {https://doi.org/10.1103/physrevapplied.14.034054} {\bibfield  {journal}
  {\bibinfo  {journal} {Phys.\ Rev.\ Appl.}\ }\textbf {\bibinfo {volume}
  {14}},\ \bibinfo {pages} {034054} (\bibinfo {year} {2020})}\BibitemShut
  {NoStop}%
\bibitem [{\citenamefont {Ding}\ \emph {et~al.}(2019)\citenamefont {Ding},
  \citenamefont {Dong}, \citenamefont {Zhang}, \citenamefont {Shi},
  \citenamefont {Yu}, \citenamefont {Ye}, \citenamefont {Guo},\ and\
  \citenamefont {Shi}}]{Ding2019}%
  \BibitemOpen
  \bibfield  {author} {\bibinfo {author} {\bibfnamefont {D.-S.}\ \bibnamefont
  {Ding}}, \bibinfo {author} {\bibfnamefont {M.-X.}\ \bibnamefont {Dong}},
  \bibinfo {author} {\bibfnamefont {W.}~\bibnamefont {Zhang}}, \bibinfo
  {author} {\bibfnamefont {S.}~\bibnamefont {Shi}}, \bibinfo {author}
  {\bibfnamefont {Y.-C.}\ \bibnamefont {Yu}}, \bibinfo {author} {\bibfnamefont
  {Y.-H.}\ \bibnamefont {Ye}}, \bibinfo {author} {\bibfnamefont {G.-C.}\
  \bibnamefont {Guo}},\ and\ \bibinfo {author} {\bibfnamefont {B.-S.}\
  \bibnamefont {Shi}},\ }\href {https://doi.org/10.1038/s42005-019-0201-1}
  {\bibfield  {journal} {\bibinfo  {journal} {Commun.\ Phys.}\ }\textbf
  {\bibinfo {volume} {2}},\ \bibinfo {pages} {100} (\bibinfo {year}
  {2019})}\BibitemShut {NoStop}%
\bibitem [{\citenamefont {Antypas}\ \emph {et~al.}(2019)\citenamefont
  {Antypas}, \citenamefont {Tretiak}, \citenamefont {Budker},\ and\
  \citenamefont {Akulshin}}]{Antypas2019}%
  \BibitemOpen
  \bibfield  {author} {\bibinfo {author} {\bibfnamefont {D.}~\bibnamefont
  {Antypas}}, \bibinfo {author} {\bibfnamefont {O.}~\bibnamefont {Tretiak}},
  \bibinfo {author} {\bibfnamefont {D.}~\bibnamefont {Budker}},\ and\ \bibinfo
  {author} {\bibfnamefont {A.}~\bibnamefont {Akulshin}},\ }\href
  {https://doi.org/10.1364/ol.44.003657} {\bibfield  {journal} {\bibinfo
  {journal} {Optics Letters}\ }\textbf {\bibinfo {volume} {44}},\ \bibinfo
  {pages} {3657} (\bibinfo {year} {2019})}\BibitemShut {NoStop}%
\bibitem [{\citenamefont {Lam}\ \emph {et~al.}(2019)\citenamefont {Lam},
  \citenamefont {Pal}, \citenamefont {Vogt}, \citenamefont {Gross},
  \citenamefont {Kiffner},\ and\ \citenamefont {Li}}]{Lam2019}%
  \BibitemOpen
  \bibfield  {author} {\bibinfo {author} {\bibfnamefont {M.}~\bibnamefont
  {Lam}}, \bibinfo {author} {\bibfnamefont {S.~B.}\ \bibnamefont {Pal}},
  \bibinfo {author} {\bibfnamefont {T.}~\bibnamefont {Vogt}}, \bibinfo {author}
  {\bibfnamefont {C.}~\bibnamefont {Gross}}, \bibinfo {author} {\bibfnamefont
  {M.}~\bibnamefont {Kiffner}},\ and\ \bibinfo {author} {\bibfnamefont
  {W.}~\bibnamefont {Li}},\ }\href {https://doi.org/10.1364/ol.44.002931}
  {\bibfield  {journal} {\bibinfo  {journal} {Optics Letters}\ }\textbf
  {\bibinfo {volume} {44}},\ \bibinfo {pages} {2931} (\bibinfo {year}
  {2019})}\BibitemShut {NoStop}%
\bibitem [{\citenamefont {Lam}\ \emph {et~al.}(2020)\citenamefont {Lam},
  \citenamefont {Pal}, \citenamefont {Vogt}, \citenamefont {Kiffner},\ and\
  \citenamefont {Li}}]{Lam2020}%
  \BibitemOpen
  \bibfield  {author} {\bibinfo {author} {\bibfnamefont {M.}~\bibnamefont
  {Lam}}, \bibinfo {author} {\bibfnamefont {S.~B.}\ \bibnamefont {Pal}},
  \bibinfo {author} {\bibfnamefont {T.}~\bibnamefont {Vogt}}, \bibinfo {author}
  {\bibfnamefont {M.}~\bibnamefont {Kiffner}},\ and\ \bibinfo {author}
  {\bibfnamefont {W.}~\bibnamefont {Li}},\ }\href
  {https://arxiv.org/abs/2012.15449} {\bibfield  {journal} {\bibinfo  {journal}
  {arXiv:2012.15449}\ } (\bibinfo {year} {2020})}\BibitemShut {NoStop}%
\bibitem [{\citenamefont {Downes}\ \emph {et~al.}(2020)\citenamefont {Downes},
  \citenamefont {MacKellar}, \citenamefont {Whiting}, \citenamefont
  {Bourgenot}, \citenamefont {Adams},\ and\ \citenamefont
  {Weatherill}}]{Downes2020}%
  \BibitemOpen
  \bibfield  {author} {\bibinfo {author} {\bibfnamefont {L.~A.}\ \bibnamefont
  {Downes}}, \bibinfo {author} {\bibfnamefont {A.~R.}\ \bibnamefont
  {MacKellar}}, \bibinfo {author} {\bibfnamefont {D.~J.}\ \bibnamefont
  {Whiting}}, \bibinfo {author} {\bibfnamefont {C.}~\bibnamefont {Bourgenot}},
  \bibinfo {author} {\bibfnamefont {C.~S.}\ \bibnamefont {Adams}},\ and\
  \bibinfo {author} {\bibfnamefont {K.~J.}\ \bibnamefont {Weatherill}},\ }\href
  {https://doi.org/10.1103/physrevx.10.011027} {\bibfield  {journal} {\bibinfo
  {journal} {Physical Review X}\ }\textbf {\bibinfo {volume} {10}},\ \bibinfo
  {pages} {011027} (\bibinfo {year} {2020})}\BibitemShut {NoStop}%
\bibitem [{dat(2020)}]{dataset}%
  \BibitemOpen
  \bibfield  {journal} {\bibinfo  {journal} {Dataset}\ }\href
  {https://doi.org/10.15129/578985d6-ff89-4be9-aa78-8710aeeac92d}
  {10.15129/578985d6-ff89-4be9-aa78-8710aeeac92d} (\bibinfo {year}
  {2020})\BibitemShut {NoStop}%
\end{thebibliography}
%

\end{document}